\documentclass[aps,pra,twocolumn,groupedaddress,nofootinbib,longbibliography]{revtex4-1}
\usepackage{amsmath, amsfonts, amssymb,amsthm}
\usepackage{graphicx,subfigure}
\usepackage{mathrsfs}
\usepackage[mathscr]{eucal}
\usepackage[T1]{fontenc}
\usepackage{physics}
\usepackage{bm}
\usepackage{subfigure}
\usepackage[bookmarks=false]{hyperref}
\usepackage{xcolor}
\usepackage{indentfirst}

\hypersetup{colorlinks=true,citecolor=blue,
linkcolor=blue,urlcolor=blue,pdfstartview=FitH,
bookmarksopen=true}

\DeclareMathAlphabet{\mathpzc}{OT1}{pzc}{m}{it}

\begin{document}
\title{Approaching Heisenberg-scalable thermometry with built-in robustness against noise}
\author{Da-Jian Zhang}
\email{zdj@sdu.edu.cn}
\affiliation{Department of Physics, Shandong University, Jinan 250100, China}
\author{D.~M.~Tong}
\email{tdm@sdu.edu.cn}
\affiliation{Department of Physics, Shandong University, Jinan 250100, China}

\date{\today}

\begin{abstract}
It is a major goal in quantum thermometry to reach a $1/N$ scaling of thermometric precision known as Heisenberg scaling but is still in its infancy to date. The main obstacle is that the resources typically required are highly
entangled states, which are very difficult to produce and extremely vulnerable to noises. Here, we propose an entanglement-free scheme of thermometry to approach Heisenberg scaling for a wide range of $N$, which has built-in robustness irrespective of the type of noise in question. Our scheme is amenable to a variety of experimental setups. Moreover, it can be used as a basic building block for promoting previous proposals of thermometry to reach Heisenberg scaling, and its applications are not limited to thermometry but can be straightforwardly extended to other metrological tasks.
\end{abstract}

\maketitle

\section{Introduction}
Accurately measuring temperature is of universal importance, underpinning many fascinating applications in material science \cite{2005Aigouy184105,2010Linden130401},
medicine and biology \cite{2009Klinkert2661,2014Schirhagl83}, and quantum thermodynamics \cite{2004Gemmer}.
The advent of quantum technologies has opened up exciting
possibilities of exploiting quantum effects to yield the quantum enhancement of thermometric precision that cannot otherwise be obtained using classical methods \cite{2018DPS,2019Mehboudi303001}. The emerging field, known as quantum thermometry nowadays, could
have a large impact on quantum platforms demanding precise temperature control, such as cold atoms, trapped ions, and superconducting circuits. This has motivated a vibrant activity on quantum thermometry over the past decade \cite{2010Stace11611,2014Sabin6436,2015Correa220405,
2016A.DePasquale12782,
2019Seah180602,2020Kulikov240501,
2020Mitchison80402,2020Latune83049,2020Xie63844,2021Rubio190402,Zhang-An}.

A major goal in quantum thermometry is to reach a $1/N$ scaling of precision known as Heisenberg scaling (HS), which represents an important quantum advantage of central interest in quantum metrology \cite{2015DemkowiczDobrzanski345,2017Degen35002,2018Braun35006}. Here, $N$ stands for the amount of physical resources which usually refers to the number of probes employed. Conventionally, HS is achieved by exploiting highly entangled states \cite{2010Stace11611,2014Sabin6436,
2020Latune83049,2020Xie63844}. However, along this line, the HS permitted in theory is typically elusive in reality, due to the vulnerability of highly entangled states to noises as well as the difficulty in producing these states. This point has been theoretically shown in Refs.~\cite{2008Ji5172,2011Escher406,2014Demkowicz-Dobrzanski250801,
2014Alipour120405} and is also reflected in the fact that less experimental progress has been made in implementing Heisenberg-scalable thermometry to date. Indeed, although the issue of how to attain HS has received much attention since the early days of quantum thermometry \cite{2010Stace11611}, the first experiment demonstrating HS was carried out only recently \cite{2019Uhlig294}. This experiment explored NOON states to reach HS and observed that the decoherence effect becomes increasingly severe as the order of NOON states increases. As such, HS was only demonstrated for small $N$, i.e., $N\leq N_\textrm{max}=9$, for which the scaling advantage is far from allowing one to beat the best possible classical methods. Besides, several no-go theorems \cite{2008Fujiwara255304,
2012Demkowicz-Dobrzanski1063,2017DemkowiczDobrzanski41009,2021Zhou10343} show that HS is forbidden to reach in the asymptotic limit $N\rightarrow\infty$ in the presence of most types of noises. For these reasons, it is crucial to find a noise-robust scheme capable of reaching HS in the regime of large $N$ \cite{Cimini} which is of interest from a practical standpoint and allowed by these theorems.

\begin{figure}
	\includegraphics[width=0.45\textwidth]{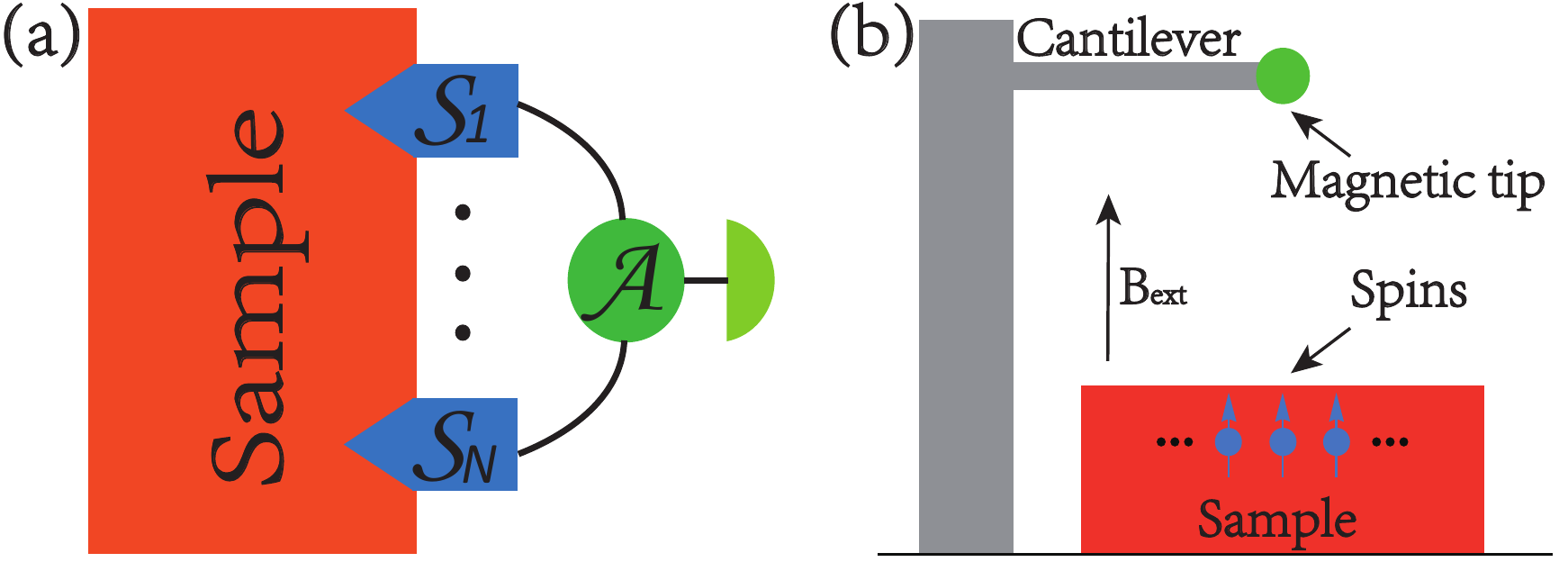}
	\caption{Schematic of our scheme. (a) Configuration of the setup implementing our scheme. To estimate the temperature of a given sample, we put $N$ probes in contact with the sample and weakly couple all of them to a same external ancilla. The continuous interplay between the strong Markovian thermalization of the probes and the relatively weak coupling to the ancilla ensures information transmission from the probes to the ancilla. This leads to the coherent accumulation of a temperature-dependent phase, and measuring the phase yields an estimate of the temperature. (b) Illustration of one possible implementation of our scheme using magnetic resonance force microscopy. The spins immersed in the sample serve as the probes and the magnetic tip serves as the ancilla. The force produced by the spins on the magnetic tip affects the mechanical vibrations of the cantilever, which can be detected by optical methods.}
	\label{fig1}
\end{figure}

Note that a number of noise-robust schemes have been proposed for other metrological tasks \cite{
2013Chaves120401,
2014Kessler150802,2014Duer80801,2014Arrad150801,2015Lu7282,2016Unden230502,2017Reiter1822,
2018Matsuzaki140501,
2018Zhou78,2019Layden40502,2019Bai40402}.
However, there has not been a noise-robust scheme in quantum thermometry so far.
In this work, we fill the gap. We find that a temperature-dependent phase can be accumulated coherently through the continuous interplay between the strong Markovian thermalization of $N$ probes and the relatively weak coupling of the $N$ probes to a same external ancilla (see Fig.~1). This mechanism enables us to propose a scheme of thermometry to approach HS for a wide range of $N$, with robustness irrespective of the type of noise in question but without complicated error correction techniques.

A salient feature of our scheme is that the whole estimation procedure is free of entanglement, unlike in previous works \cite{2018Braun35006}, where highly entangled states are either introduced in the state-preparation stage or generated via some interactions in the interrogation stage. Another feature of our scheme is that the probing time is not increasingly long with $N$. These two features distinguish our scheme from previous schemes like the parallel scheme and sequential scheme \cite{2006Giovannetti10401} on a fundamental level.  As detailed below, our scheme is amenable to a variety of experimental setups. Moreover, it can be used as a basic building block for promoting previous proposals of thermometry to reach HS, and its applications are not limited to thermometry but can be straightforwardly extended to other metrological tasks.

\section{Results}
\textbf{The basic dynamical equation.}
Suppose that we are given a sample of temperature $T$, with which a probe $\mathpzc{S}$ keeps in contact. The thermalization process of $\mathpzc{S}$ can be described by a Markovian master equation  $\partial_t\rho_\mathpzc{S}(t)=\mathcal{L}_\mathpzc{S}\rho_\mathpzc{S}(t)$ \cite{2018DPS,2019Mehboudi303001}. Here, $\rho_\mathpzc{S}(t)$ is the evolving state of $\mathpzc{S}$. $\mathcal{L}_\mathpzc{S}$ is a Liouville superoperator assumed to have the following two properties: (i) it admits  a temperature-dependent state $\rho_{_T}$ as the unique steady state; and (ii) the nonzero eigenvalues $\lambda_\mu$ of $\mathcal{L}_\mathpzc{S}$ have negative real parts, where $\mu$ is an index labeling the eigenvalues. Property (i) implies that $\mathcal{L}_\mathpzc{S}\rho_{_T}=0$, and property (ii) means that there is a dissipative gap $\lambda:={\min_{\lambda_\mu\neq 0}\abs{\Re(\lambda_\mu)}}$ in the Liouvillian spectrum \cite{2016Zhang12117,2016Zhang52132,2020Zhang23418}. We do not impose any
restriction on the explicit forms of $\mathcal{L}_\mathpzc{S}$ and $\rho_{_T}$, so that the scheme to be presented is applicable to a variety of physical models. In particular, $\rho_{_T}$ may or may not be the Gibbs state, depending on the specific physical model in question. Starting from an arbitrary initial state, $\rho_\mathpzc{S}(t)$ undergoing the thermalization process automatically evolves towards $\rho_{_T}$ at an exponentially fast rate \cite{2016Zhang12117,2016Zhang52132,2020Zhang23418}.
Therefore, we may assume that $\mathpzc{S}$ (approximately) reaches $\rho_{_T}$ at a certain time $t_0$.

We weakly couple $\mathpzc{S}$ to an ancilla $\mathpzc{A}$ after the time $t_0$. The free Hamiltonian of $\mathpzc{A}$ is denoted by $H_\mathpzc{A}$. The interaction between $\mathpzc{S}$ and $\mathpzc{A}$ is assumed to be of the form $H_I=\hbar g S\otimes A$ with $[A, H_\mathpzc{A}]=0$, where $g$ is a positive number describing the coupling strength, and $S$ and $A$ are two Hermitian operators of $\mathpzc{S}$ and $\mathpzc{A}$, respectively. In the frame rotating with $H_\mathpzc{A}$, the dynamics of $\mathpzc{S}\mathpzc{A}$ can be described by
\begin{eqnarray}\label{eq:main}
\partial_t\rho(t)=\mathcal{L}_\mathpzc{S}\rho(t)-\textrm{i}g[S\otimes A,\rho(t)]=:\mathcal{L}\rho(t).
\end{eqnarray}
Note that we temporarily do not take noise into account, as our purpose here is to show how the continuous interplay between the thermalization $\mathcal{L}_\mathpzc{S}$ and the interaction $H_I$ leads to the coherent accumulation of a temperature-dependent phase.

Let us figure out the reduced dynamics of $\mathpzc{A}$. Inspired by the Nakajima-Zwanzig projection operator technique \cite{2007BreuerPetruccione}, we introduce a superoperator $\mathcal{P}$, defined as $\mathcal{P}X=\rho_{_T}\otimes\tr_\mathpzc{S}X$, for an operator $X$ acting on the joint Hilbert space of $\mathpzc{SA}$. Evidently, $\mathcal{P}\rho(t)=\rho_{_T}\otimes\rho_\mathpzc{A}(t)$, where $\rho_\mathpzc{A}(t)$ denotes the evolving state of $\mathpzc{A}$. The superoperator complementary to $\mathcal{P}$, denoted by $\mathcal{Q}$, is defined as $\mathcal{Q}X=X-\mathcal{P}X$.

Applying $\mathcal{P}$, $\mathcal{Q}$ to Eq.~(\ref{eq:main}) separately and invoking $\mathcal{P}+\mathcal{Q}=1$, where $1$ stands for the identity map, we have
\begin{eqnarray}\label{eq:relevant}
\partial_t\mathcal{P}\rho(t)=\mathcal{P}\mathcal{L}\mathcal{P}\rho(t)+
\mathcal{P}\mathcal{L}\mathcal{Q}\rho(t),
\end{eqnarray}
\begin{eqnarray}\label{eq:irrelavant}
\partial_t\mathcal{Q}\rho(t)=\mathcal{Q}\mathcal{L}\mathcal{P}\rho(t)+
\mathcal{Q}\mathcal{L}\mathcal{Q}\rho(t).
\end{eqnarray}
The solution of Eq.~(\ref{eq:irrelavant}) can be formally expressed as \cite{2007BreuerPetruccione}
\begin{eqnarray}\label{eq:formal-solution}
\mathcal{Q}\rho(t)=\mathcal{G}(t,t_0)\mathcal{Q}\rho(t_0)+\int_{t_0}^t ds\mathcal{G}(t,s)\mathcal{Q}\mathcal{L}\mathcal{P}\rho(s),
\end{eqnarray}
where $\mathcal{G}(t,s):=e^{\mathcal{Q}\mathcal{L}(t-s)}$. This point can be verified by inserting Eq.~(\ref{eq:formal-solution}) into Eq.~(\ref{eq:irrelavant}).
Substituting Eq.~(\ref{eq:formal-solution}) into Eq.~(\ref{eq:relevant}) and noting that $\rho(t_0)=\rho_{_T}\otimes\rho_\mathpzc{A}(t_0)$ and therefore $\mathcal{Q}\rho(t_0)=0$, we obtain an exact equation of motion for $\mathpzc{A}$:
\begin{eqnarray}\label{eq:relevant-closed}
\partial_t\mathcal{P}\rho(t)=\mathcal{P}\mathcal{L}\mathcal{P}\rho(t)+
\int_{t_0}^t ds\mathcal{P}\mathcal{L}\mathcal{G}(t,s)\mathcal{Q}\mathcal{L}\mathcal{P}\rho(s).
\end{eqnarray}
Here, the second term on the right-hand side represents some memory effect, as it depends on the past history of $\mathcal{P}\rho(s)$.

To figure out the explicit expression for the first term on the right-hand side of Eq.~(\ref{eq:relevant-closed}), we deduce from property (i) that $\mathcal{L}_\mathpzc{S}\rho_{_T}=0$ and hence $\mathcal{LP}\rho(t)=-\textrm{i}g[S\otimes A,\rho_{_T}\otimes\rho_\mathpzc{A}(t)]$. Then, from the defining property of $\mathcal{P}$, it follows that
\begin{eqnarray}\label{eq:first-term}
\mathcal{P}\mathcal{L}\mathcal{P}\rho(t)=\rho_{_T}\otimes
\left(-\textrm{i}g[\varphi_{_T} A,\rho_\mathpzc{A}(t)]\right),
\end{eqnarray}
where
$\varphi_{_T}:=\tr(S\rho_{_T})$. On the other hand, using property (ii), the norm of the second term on the right-hand side of Eq.~(\ref{eq:relevant-closed}) can be bounded as
\begin{eqnarray}\label{bound-memory-effect}
\norm{\int_{t_0}^t ds\mathcal{P}\mathcal{L}
\mathcal{G}(t,s)\mathcal{Q}\mathcal{L}\mathcal{P}\rho(s)}
\leq\frac{\varepsilon g\norm{\mathcal{P}}^2\norm{\mathcal{K}}^2}{(\lambda/g)-\varepsilon\norm{\mathcal{K}}},
\end{eqnarray}
where $\mathcal{K}$ is the superoperator defined as $\mathcal{K}X:=-\textrm{i}[S\otimes A, X]$, and
$\varepsilon$ is a dimensionless constant determined by the damping basis
of $\mathcal{L}_\mathpzc{S}$ \cite{2016Zhang12117,2016Zhang52132}. The proof of the bound is given in the Method section.

Note that $\lambda$ characterizes the strength of the thermalization $\mathcal{L}_\mathpzc{S}$; that is, the larger $\lambda$ is, the stronger the thermalization is. Likewise, $g$ characterizes the strength of the interaction $H_I$. Under the condition that the thermalization is strong whereas the interaction is relatively weak, i.e., $\lambda/g\gg 1$, we deduce from Eq.~(\ref{bound-memory-effect}) that the second term on the right-hand side of Eq.~(\ref{eq:relevant-closed}), i.e., the memory effect, is negligible. Upon neglecting this term and inserting Eq.~(\ref{eq:first-term}) into Eq.~(\ref{eq:relevant-closed}), we reach the basic dynamical equation:
\begin{eqnarray}\label{key-equation}
\partial_t\rho_\mathpzc{A}(t)=-\textrm{i}g[\varphi_{_T} A,\rho_\mathpzc{A}(t)],
\end{eqnarray}
representing a unitary dynamics able to coherently accumulate $\varphi_{_T}$ in the state of $\mathpzc{A}$ superposed by eigenstates of $A$. Hereafter, in line with the studies on quantum phase estimation \cite{2006Giovannetti10401}, we refer to $\varphi_{_T}$ as a temperature-dependent phase.

\vspace{1em}
\textbf{The scheme of thermometry.}
To estimate the unknown temperature $T$ of a given sample,
we put $N$ probes, $\mathpzc{S}_1,\cdots,\mathpzc{S}_N$, in contact with the sample. Upon preparing each probe in $\rho_{_T}$ at time $t_0$, we couple all of the probes to a same ancilla $\mathpzc{A}$. The above configuration is schematically shown in Fig.~1a. A natural platform for implementing it could be the magnetic resonance force microscopy \cite{2006Berman} (see Fig.~1b), where the spins immersed in the sample and the magnetic tip serve as the probes and the ancilla, respectively. The dynamical equation for the $N$ probes and the ancilla in the rotating frame reads
\begin{eqnarray}\label{scheme-Lindblad-equation}
\partial_t\rho(t)=\left(\mathcal{L}_{\mathpzc{S}_1\mathpzc{A}}+\cdots+
\mathcal{L}_{\mathpzc{S}_N\mathpzc{A}}\right)\rho(t),
\end{eqnarray}
with $\mathcal{L}_{\mathpzc{S}_n\mathpzc{A}}\rho=\mathcal{L}_{\mathpzc{S}_n}\rho- \textrm{i}g_n[S_n\otimes A,\rho]$. Here, $\mathcal{L}_{\mathpzc{S}_n}$ denotes the Liouville superoperator for $\mathpzc{S}_n$, $g_n$ represents the coupling strength between $\mathpzc{S}_n$ and $\mathpzc{A}$, and $S_n$ is a Hermitian operator acting on $\mathpzc{S}_n$. For simplicity, we assume that $\mathcal{L}_{\mathpzc{S}_n}=\mathcal{L}_{\mathpzc{S}}$, $g_n=g$, and $S_n=S$, for all $n$. Our following discussion can be straightforwardly extended to the general case.

Let $\mathcal{E}_N(t_1,t_0)$ denote the dynamical map associated with Eq.~(\ref{scheme-Lindblad-equation}),
\begin{eqnarray}\label{scheme-DP}
\mathcal{E}_N(t_1,t_0)=\exp\left[\left(\mathcal{L}_{\mathpzc{S}_1\mathpzc{A}}+\cdots+
\mathcal{L}_{\mathpzc{S}_N\mathpzc{A}}\right)(t_1-t_0)\right],
\end{eqnarray}
transforming the state $\rho_{_T}^{\otimes N}\otimes\rho_\mathpzc{A}(t_0)$ at time $t_0$ into the state $\mathcal{E}_N(t_1,t_0)\rho_{_T}^{\otimes N}\otimes\rho_\mathpzc{A}(t_0)$ at time $t_1$. Then, the state of $\mathpzc{A}$ at time $t_1$ reads
\begin{eqnarray}\label{output-state-A}
\rho_\mathpzc{A}(t_1)=\tr_{\mathpzc{S}_1\cdots\mathpzc{S}_N}
\left[\mathcal{E}_N(t_1,t_0)\rho_{_T}^{\otimes N}\otimes\rho_\mathpzc{A}(t_0)\right].
\end{eqnarray}
To find the explicit expression for $\rho_\mathpzc{A}(t_1)$, we resort to the fact that $\mathcal{L}_{\mathpzc{S}_n\mathpzc{A}}$'s commute with each other and rewrite Eq.~(\ref{scheme-DP}) as
\begin{eqnarray}\label{DP-decomposition}
\mathcal{E}_N(t_1,t_0)=\mathcal{E}_{\mathpzc{S}_1\mathpzc{A}}(t_1,t_0)\cdots
\mathcal{E}_{\mathpzc{S}_N\mathpzc{A}}(t_1,t_0),
\end{eqnarray}
with $\mathcal{E}_{\mathpzc{S}_n\mathpzc{A}}(t_1,t_0)=\exp[\mathcal{L}_{\mathpzc{S}_n\mathpzc{A}}
(t_1-t_0)]$. Besides, according to the basic dynamical equation (\ref{key-equation}),
\begin{eqnarray}\label{UD-single}
&&\tr_{\mathpzc{S}_n}\left[\mathcal{E}_{\mathpzc{S}_n\mathpzc{A}}(t_1,t_0)\rho_{_T}\otimes
\rho_\mathpzc{A}(t_0)\right]
=\exp[-\textrm{i}g\varphi_{_T}A(t_1-t_0)]\times\nonumber\\
&&\rho_\mathpzc{A}(t_0)\exp[\textrm{i}g\varphi_{_T}A(t_1-t_0)].
\end{eqnarray}
Substituting Eqs.~(\ref{DP-decomposition}) and (\ref{UD-single}) into Eq.~(\ref{output-state-A}), we have
\begin{eqnarray}\label{output-state-ancilla}
\rho_\mathpzc{A}(t_1)=e^{-\textrm{i}gN\varphi_{_T}A(t_1-t_0)}
\rho_\mathpzc{A}(t_0)e^{\textrm{i}gN\varphi_{_T}A(t_1-t_0)},
\end{eqnarray}
implying that $\varphi_{_T}$ is intensified $N$ times due to the use of $N$ probes.

In deriving Eq.~(\ref{UD-single}), we have neglected the memory effect. Indeed, the memory effect can be made arbitrarily weak by choosing either a large $\lambda$ or a small $g$ so that $\lambda/g$ is large enough. Here, $N$ should be confined to a certain appropriate range, as detailed below. It is worth noting that the past two decades have witnessed enormous experimental progresses in engineering strongly dissipative processes (see the review article \cite{2012Mueller1} and references therein). Nowadays, experimentalists are able to realize arbitrary Markovian processes for a number of experimentally mature platforms such as trapped ions \cite{2013Schindler123012} and Rydberg atoms \cite{2010Weimer382}. These developments imply that it is experimentally feasible to obtain a large $\lambda$. Besides, the value of $g$ is often fixed once the platform adopted has been calibrated, whereas the value of $\lambda$ may be still allowed to be tuned. For example, in the magnetic resonance force microscopy, the value of $g$ is determined by the distance between the spins and the magnetic tip \cite{2006Berman}. So, $g$ is fixed once the distance has been fixed. However, $\lambda$ is still tunable through varying the intensity and frequency of an external magnetic field \cite{2006Berman}.  For the above reasons, we consider the scenario that $\lambda$ is large whereas $g$ is fixed.

Without loss of generality, we set the probing time $t_1-t_0$ to be a ``unit'' of time $1/g$ for simplicity. It follows from Eq.~(\ref{output-state-ancilla}) that
\begin{eqnarray}
\rho_\mathpzc{A}(t_1)=\exp[-\textrm{i}N\varphi_{_T}A]
\rho_\mathpzc{A}(t_0)\exp[\textrm{i}N\varphi_{_T}A].
\end{eqnarray}
Then, an estimate of $T$ can be obtained by measuring $\rho_\mathpzc{A}(t_1)$.
It remains to find the optimal state $\rho_\mathpzc{A}(t_0)$ and observable $O$. This can be carried out with the aid of
the quantum Cram\'{e}r-Rao theorem \cite{1994Braunstein3439}, stating that the estimation error $\delta T$ from measuring any observable is bounded by the inequality $\delta T\geq1/\sqrt{\nu \mathcal{F}_N(T)}$, where $\nu$ denotes the number of measurements, and $\mathcal{F}_N(T)$ is the quantum Fisher information (QFI) for $\rho_\mathpzc{A}(t_1)$. Therefore, the optimal $\rho_\mathpzc{A}(t_0)$ is the state maximizing the QFI, and the optimal $O$ is the observable saturating the inequality. It is well-known \cite{2006Giovannetti10401} that such $\rho_\mathpzc{A}(t_0)$ and $O$ are respectively $\left(\ket{a_{M}}+\ket{a_{m}}\right)/\sqrt{2}$ and $\ket{a_M}\bra{a_m}+\ket{a_m}\bra{a_M}$. Here, $\ket{a_M}$ and $\ket{a_m}$ denote the eigenvectors of $A$ corresponding to the maximum eigenvalue $a_M$ and minimum eigenvalue $a_m$, respectively.

In light of the above analysis, we may now specify our scheme as follows: (i) Put $N$ probes $\mathpzc{S}_1,\cdots,\mathpzc{S}_N$ in contact with the sample and prepare each of them in $\rho_{_T}$ at time $t_0$. (ii) Initialize the state of ancilla $\mathpzc{A}$ to be $\left(\ket{a_M}+\ket{a_m}\right)/\sqrt{2}$ and couple all of the probes to $\mathpzc{A}$ for the time $1/g$. (iii) Perform the measurement associated with $O=\ket{a_M}\bra{a_m}+\ket{a_m}\bra{a_M}$ on $\mathpzc{A}$ at time $t_1=t_0+1/g$. (iv) Repeat steps (ii) and (iii) $\nu$ times to build a sufficient statistics for determining $T$.
As the QFI for $\rho_\mathpzc{A}(t_1)$ reads
\begin{eqnarray}\label{QFI}
\mathcal{F}_N(T)=N^2(a_M-a_m)^2\abs{\partial_{_T}\varphi_{_T}}^2,
\end{eqnarray}
the precision attained in our scheme approaches the HS
\begin{eqnarray}\label{sensitivity}
\delta T={1}/\left[{\sqrt{\nu} N(a_M-a_m)\abs{\partial_{_T}\varphi_{_T}}}\right].
\end{eqnarray}
Notably, the evolving state of the $N$ probes and the ancilla is approximately $\rho_{_T}^{\otimes N}\otimes \rho_\mathpzc{A}(t)$. Thus, no highly entangled state is involved in the whole estimation procedure of our scheme. Moreover, the probing time in our scheme is a single ``unit'' $1/g$ and does not increase as $N$ increases. In passing, we point out that the precision of the form (\ref{sensitivity}) is termed as the weak Heisenberg limit in Ref.~\cite{2017Luis32113}.

It may be instructive to give a physical picture for comprehending how our scheme works.
To gain some insight into the physical origin of Eq.~(\ref{key-equation}), we may interpret the probe employed in our scheme as an effective transducer, which helps to extract the information about temperature from the sample and transmit it to the ancilla under the condition that the Markovian thermalization is strong enough so that the memory effect is negligible. The ancilla, on the other hand, may be interpreted as an information storage and is responsible for storing the information about temperature, with its initial state determining the storage capacity. Such a continuous interplay between the probe and the ancilla leads to the basic dynamical equation (\ref{key-equation}) permitting the coherent accumulation of the temperature-dependent phase $\varphi_{_T}$ in the state of the ancilla. The accumulating rate is described by $g$, which is reminiscent of the transduction parameter in electrical or magnetic field sensing \cite{2017Degen35002}. When $N$ probes are coupled to the same ancilla, the accumulating rate is increased from $g$ to $Ng$ as indicated in Eq.~(\ref{output-state-ancilla}), since there are now $N$ effective transducers transmitting the information about temperature to the ancilla. We emphasize that the working principle of our scheme is built upon the very nature of open dynamics, which distinguishes our scheme from previous schemes involving Ramsey and Mech-Zehnder interferometers where the dynamics is unitary.
	
\vspace{1em}
\textbf{Built-in robustness of our scheme against noises.}
Taking noises into account, we shall rewrite Eq.~(\ref{scheme-Lindblad-equation}) as
\begin{eqnarray}\label{full-dynamics-noise}
\partial_t\rho(t)=&&
[(\mathcal{L}_{\mathpzc{S}_1\mathpzc{A}}+
\mathcal{L}_{\mathpzc{S}_1}^\textrm{noise})
+\cdots+
(\mathcal{L}_{\mathpzc{S}_N\mathpzc{A}}+\mathcal{L}_{\mathpzc{S}_N}^\textrm{noise}
)+\nonumber\\
&&\mathcal{L}_{\mathpzc{A}}^\textrm{noise}]\rho(t),
\end{eqnarray}
with $\mathcal{L}_{\mathpzc{S}_n}^{\textrm{noise}}$ and $\mathcal{L}_{\mathpzc{A}}^{\textrm{noise}}$ denoting the Liouville superoperators describing the noises acting on $\mathpzc{S}_n$ and $\mathpzc{A}$, respectively. Here, no restriction is imposed on the explicit forms of $\mathcal{L}_{\mathpzc{S}_n}^{\textrm{noise}}$ and $\mathcal{L}_{\mathpzc{A}}^{\textrm{noise}}$. Under the condition that the Markovian thermalization $\mathcal{L}_{\mathpzc{S}_n}$ is strong, the QFI of our scheme in the presence of the noises can be evaluated as
\begin{eqnarray}\label{OS-QFI-general-main}
\mathcal{F}_N^\textrm{noise}(T)=2N^2 \abs{\partial_{_T}\varphi_{_T}}^2\sum_{k\neq l}p_{k,l}\abs{\bra{\phi_k}A\ket{\phi_l}}^2,
\end{eqnarray}
with coefficients
\begin{eqnarray}
p_{k,l}=
\begin{cases}
  0, & \mbox{if } p_k=p_l=0, \\
  \frac{(p_k-p_l)^2}{p_k+p_l}, & \mbox{otherwise}.
\end{cases}
\end{eqnarray}
Here, $p_k$ and $\ket{\phi_k}$ are the eigenvalue and associated eigenvector of a density operator $\delta$ which is determined by $\mathcal{L}_\mathpzc{A}^\textrm{noise}$ and $\rho_\mathpzc{A}(t_0)$. The proof of Eq.~(\ref{OS-QFI-general-main}) is presented in Supplementary Note 1, where the expression of $\delta$ is given. Using Eqs.~(\ref{QFI}) and (\ref{OS-QFI-general-main}), we arrive at the result that the ratio ${\mathcal{F}_N^\textrm{noise}(T)}/{\mathcal{F}_N(T)}$ is independent of $N$. This result means that the detrimental influence of noises on our scheme does not become increasingly severe as $N$ increases.

To see the significance of the above result, we may recall the reason why the HS permitted in many previous schemes is extremely vulnerable to noises. So far, a lot of effort has been devoted to exploring the possibilities of encoding the parameter of interest into the relative
phase of a quantum system for reaching HS, which has become one main stream of research on Heisenberg-scalable metrology nowadays. Generally speaking, along this line of development, previous schemes require either highly entangled states or increasingly long probing time. A well-known example is the parallel scheme \cite{2006Giovannetti10401}, in which $N$ probes are employed and a relative phase $N\varphi$ is obtained by preparing the initial state of the probes to be an $N$-body maximally entangled state. Evidently, the $N$-body maximally entangled state is increasingly vulnerable to noises as $N$ increases. Hence, in the presence of noises, the parallel scheme usually suffers from an exponential drop in performance as $N$ increases \cite{2017Degen35002,2018Braun35006}. Another well-known example is the sequential scheme \cite{2006Giovannetti10401}, in which a single probe is employed and a relative phase $N\varphi$  is obtained with an $N$ times long probing time. Since the decoherence effect is accumulated with time, the longer the probing time is, the severer the detrimental influence of noises on the sequential scheme becomes. Likewise, the performance of the sequential scheme usually drops exponentially as $N$ increases in the presence of noises \cite{2017Degen35002,2018Braun35006}.
As a matter of fact, most of the experiments relying on highly entangled states or increasingly long probing time are confined to the regime of very small $N$ \cite{2017Degen35002,2018Braun35006}.

Our scheme provides a {different means} of encoding the parameter of interest into the relative
phase of a quantum system without appealing to highly entangled states and increasingly long probing time. Here, by saying ``{different means},'' we mean that the phase $N\varphi_{_T}$ stems from the information transmission from the $N$ probes to the same ancilla, which are permitted by the very nature of open dynamics. The key advantage of exploring
this {means} for reaching HS is that the detrimental influence of noises is no longer increasingly severe as $N$ increases. This means that our scheme is robust against noises, which, as demonstrated below, allows for approaching HS for a wide range of $N$. Notably, the robustness of our scheme is intrinsic, since no error correction technique is involved in our scheme. Moreover, it is worth noting that the robustness of our scheme is irrespective of the specific type of noises in question, since Eq.~(\ref{OS-QFI-general-main}) is derived without restricting the explicit forms of $\mathcal{L}_{\mathpzc{S}_n}^{\textrm{noise}}$ and $\mathcal{L}_{\mathpzc{A}}^{\textrm{noise}}$. In passing, we would like to point out that it has been shown in several works \cite{2010Stace11611,2014Sabin6436} that the problem of temperature estimation can be mapped into the problem of phase estimation. Analogous to the parallel scheme, the schemes in these works are based on highly entangled states like NOON states and have been experimentally shown to be very vulnerable to noises \cite{2019Uhlig294}.

\vspace{1em}
\textbf{Example.}
Let us furnish an analytically
solvable model to demonstrate the usefulness of our scheme. Consider the physical model of $N$ qubits independently contacting with a Bosonic thermal reservoir of temperature $T$ \cite{2019Seah180602}. We take the qubits to be $\mathpzc{S}_1,\cdots,\mathpzc{S}_N$. The thermalization process of $\mathpzc{S}_n$ is governed by the Liouvillian
\begin{eqnarray}
\mathcal{L}_{\mathpzc{S}_n}\rho=-\frac{\textrm{i}}{\hbar}[H_{\mathpzc{S}_n},\rho]+\gamma
(\overline{n}+1)\mathcal{D}[\sigma_-^{\mathpzc{S}_n}]\rho+\gamma
\overline{n}\mathcal{D}[\sigma_+^{\mathpzc{S}_n}]\rho.\nonumber\\
\end{eqnarray}
Here, $H_{\mathpzc{S}_n}=\hbar\Omega\sigma_z^{\mathpzc{S}_n}/2$ is the free Hamiltonian of $\mathpzc{S}_n$, with $\sigma_i^{\mathpzc{S}_n}$, $i=x,y,z$, denoting the Pauli matrices acting on ${\mathpzc{S}_n}$. The last two terms stand for the process of energy exchange with the reservoir, with $\gamma>0$ describing the exchange rate. Note that the dissipative gap $\lambda$ is proportional to $\gamma$, i.e., $\lambda\propto\gamma$. $\overline{n}=[\exp\left(\hbar\Omega/k_B T\right)-1]^{-1}$ and $\mathcal{D}[\sigma_\pm^{\mathpzc{S}_n}]\rho=\sigma_\pm^{\mathpzc{S}_n}\rho\sigma_\mp^{\mathpzc{S}_n}
-\{\sigma_\mp^{\mathpzc{S}_n} \sigma_\pm^{\mathpzc{S}_n},\rho\}/2$, with $\sigma_\pm^{\mathpzc{S}_n}=(\sigma_x^{\mathpzc{S}_n}\mp i\sigma_y^{\mathpzc{S}_n})/2$. Note that the unique steady state of $\mathpzc{S}_n$ happens to be the Gibbs state $\rho_{_T}=\exp(-H_{\mathpzc{S}_n}/k_B T)/Z_T$, where $k_B$ is the Boltzmann's constant and $Z_T=\tr\exp(-H_{\mathpzc{S}_n}/k_B T)$ the partition function. $\mathpzc{A}$ is also taken to be a qubit. The interaction between $\mathpzc{S}_n$ and $\mathpzc{A}$ is set to be $\hbar g\sigma_z^{\mathpzc{S}_n}\otimes\sigma_z^\mathpzc{A}$. So,
\begin{eqnarray}
\mathcal{L}_{\mathpzc{S}_n\mathpzc{A}}\rho=\mathcal{L}_{\mathpzc{S}_n}\rho
-\textrm{i}g[\sigma_z^{\mathpzc{S}_n}\otimes \sigma_z^\mathpzc{A},\rho].
\end{eqnarray}
We examine a major
source of noises, dephasing. That is, the Liouville superoperator $\mathcal{L}_{\mathpzc{S}_n/\mathpzc{A}}^\textrm{noise}$ describing the noise acting on $\mathpzc{S}_n/\mathpzc{A}$ reads
\begin{eqnarray}
\mathcal{L}_{\mathpzc{S}_n/\mathpzc{A}}^\textrm{noise}\rho
=\kappa_{\mathpzc{S}_n/\mathpzc{A}}\left(\sigma_z^{\mathpzc{S}_n/\mathpzc{A}}
\rho\sigma_z^{\mathpzc{S}_n/\mathpzc{A}}-\rho\right),
\end{eqnarray}
where $\kappa_{\mathpzc{S}_n/\mathpzc{A}}$ denotes the dephasing rate. The initial state of the probes and the ancilla is set to be $\rho_{_T}^{\otimes N}\otimes\rho_{\mathpzc{A}}(t_0)$ with $\rho_{\mathpzc{A}}(t_0)=(\ket{0}+\ket{1})
(\bra{0}+\bra{1})/2$. So far, we have specified all the Liouville superoperators entering the full dynamical equation (\ref{full-dynamics-noise}) as well as the initial state assumed in our scheme.

Now, let us figure out the performance of our scheme in the presence of the noises. To this end, we need to find out the output state of $\mathpzc{A}$ at time $t_1=t_0+1/g$, which is given by
\begin{eqnarray}\label{Example-output-state}
\rho_\mathpzc{A}(t_1)=
\frac{1}{2}
\begin{pmatrix}
  1 & \Gamma^N e^{-2\eta} \\
  \Gamma^{*N} e^{-2\eta} & 1
\end{pmatrix}.
\end{eqnarray}
Here, $1/g$ is the probing time which is independent of $N$, and
\begin{eqnarray}\label{Example-Gamma}
\Gamma=&&\frac{e^{\omega_+(\xi)}}{2}
\left[1+\frac{(2\overline{n}+1)\xi}{\sqrt{\Delta(\xi)}}-
\frac{4\textrm{i}}{(2\overline{n}+1)\sqrt{\Delta(\xi)}}\right]
+\nonumber\\
&&\frac{e^{\omega_-(\xi)}}{2}
\left[1-\frac{(2\overline{n}+1)\xi}{\sqrt{\Delta(\xi)}}+
\frac{4\textrm{i}}{(2\overline{n}+1)\sqrt{\Delta(\xi)}}\right],
\end{eqnarray}
where $\Delta(\xi)=(2\overline{n}+1)^2\xi^2-8\textrm{i}\xi-16$ and $\omega_\pm(\xi)=[-(2\overline{n}+1)\xi\pm\sqrt{\Delta(\xi)}]/2$, with $\xi=\gamma/g$ and $\eta=\kappa_\mathpzc{A}/g$ characterizing (in units of $g$) the strength of the Markovian thermalization and that of the dephasing noise acting on the ancilla $\mathpzc{A}$, respectively. The proof of Eq.~(\ref{Example-output-state}) is presented in Supplementary Note 2. It is interesting to note that the noises acting on the probes $\mathpzc{S}_1,\cdots,\mathpzc{S}_N$ do not affect the output state $\rho_\mathpzc{A}(t_1)$. The QFI for $\rho_\mathpzc{A}(t_1)$ reads
\begin{eqnarray}\label{Example-QFI}
\mathcal{F}_N^\textrm{noise}(T)=&&
N^2\abs{\Gamma}^{2N-2}\abs{\partial_{_T}\Gamma}^2e^{-4\eta}+
N^2\abs{\Gamma}^{4N-2}\times\nonumber\\
&&\left(\partial_{_T}\abs{\Gamma}\right)^2e^{-8\eta}
/\left(1-\abs{\Gamma}^{2N}e^{-4\eta}\right).
\end{eqnarray}
Equations (\ref{Example-output-state}) and (\ref{Example-QFI}) are analytical results without numerical approximation.

We first examine the limiting case of $\xi\rightarrow\infty$, which corresponds to the ideal situation that the Markovian thermalization is infinitely strong and there is no memory effect. Using Eq.~(\ref{Example-QFI}) and
noting that $\lim_{\xi\rightarrow\infty}\Gamma=\exp(-2\textrm{i}\varphi_{_T})$, where $\varphi_{_T}:=\tr(\sigma_z\rho_{_T})=1/(2\overline{n}+1)$, we have that
\begin{eqnarray}\label{Example-QFI-ideal}
\mathcal{F}_N^\textrm{noise}(T)=
4N^2\abs{\partial_{_T}\varphi_{_T}}^2 e^{-4\eta}.
\end{eqnarray}
Equation (\ref{Example-QFI-ideal}) represents a HS of precision. Since
\begin{eqnarray}\label{Example-output-state-ideal}
\rho_\mathpzc{A}(t_1)=
\frac{1}{2}
\begin{pmatrix}
  1 & e^{-2\textrm{i}N\varphi_{_T}} e^{-2\eta} \\
  e^{2\textrm{i}N\varphi_{_T}} e^{-2\eta} & 1
\end{pmatrix},
\end{eqnarray}
we see that the HS in Eq.~(\ref{Example-QFI-ideal}) stems from the relative phase $N\varphi_{_T}$ acquired in $\rho_\mathpzc{A}(t_1)$. Notably, this relative phase is obtained without appealing to highly entangled states and increasingly long probing time. A consequence of this fact is that the detrimental influence of noises does not lead to an exponential drop of the HS. Indeed, using Eq.~(\ref{Example-QFI-ideal}) and noting that $\mathcal{F}_N(T)=4N^2\abs{\partial_{_T}\varphi_{_T}}^2$, we have that
\begin{eqnarray}\label{Example-rotio}
\mathcal{F}_N^\textrm{noise}(T)/\mathcal{F}_N(T)=e^{-4\eta},
\end{eqnarray}
which is consistent with our general analysis that $\mathcal{F}_N^\textrm{noise}(T)/\mathcal{F}_N(T)$ is independent of $N$.

We now examine the situation that the strength of the Markovian thermalization is large and within current experimental reach. To do this, we numerically compute $\mathcal{F}_N^\textrm{noise}(T)$ in units of $\mathcal{F}_\textrm{th}(T)$ for different $\xi$. Here,  $\mathcal{F}_\textrm{th}(T)=(\hbar\Omega/2k_BT^2)^2\sech^2(\hbar\Omega/2k_BT)$ is the QFI for the Gibbs state $\rho_{_T}=\exp(-H_{\mathpzc{S}_n}/k_B T)/Z_T$ \cite{2019Seah180602}. We choose four values of $\xi$, namely, $100$, $200$, $300$, and $400$. To reach these values, we can set, for instance, $g=100$ kHz and $\gamma=10$, $20$, $30$, $40$ MHz, which are experimentally feasible for a number of  quantum platforms such as Rydberg atoms and trapped ions \cite{2012Mueller1}. We examine
an exemplary temperature satisfying $k_B T/\hbar\Omega=2$, which is of interest in quantum thermometry \cite{2019Seah180602}. The numerical results are shown in Fig.~2, where $\eta$ is set to be $1/10$ in view of the convention that the lifetime of an ancilla is usually assumed to be long in quantum metrology (see, e.g., Refs.~\cite{
2014Kessler150802,2014Duer80801,2014Arrad150801,2016Unden230502,
2018Zhou78}). Here, we also show the QFI given by Eq.~(\ref{Example-QFI-ideal}) (see the black solid line), which serves as a benchmark for the QFIs associated with the four values of $\xi$.

\begin{figure}
	\includegraphics[width=0.45\textwidth]{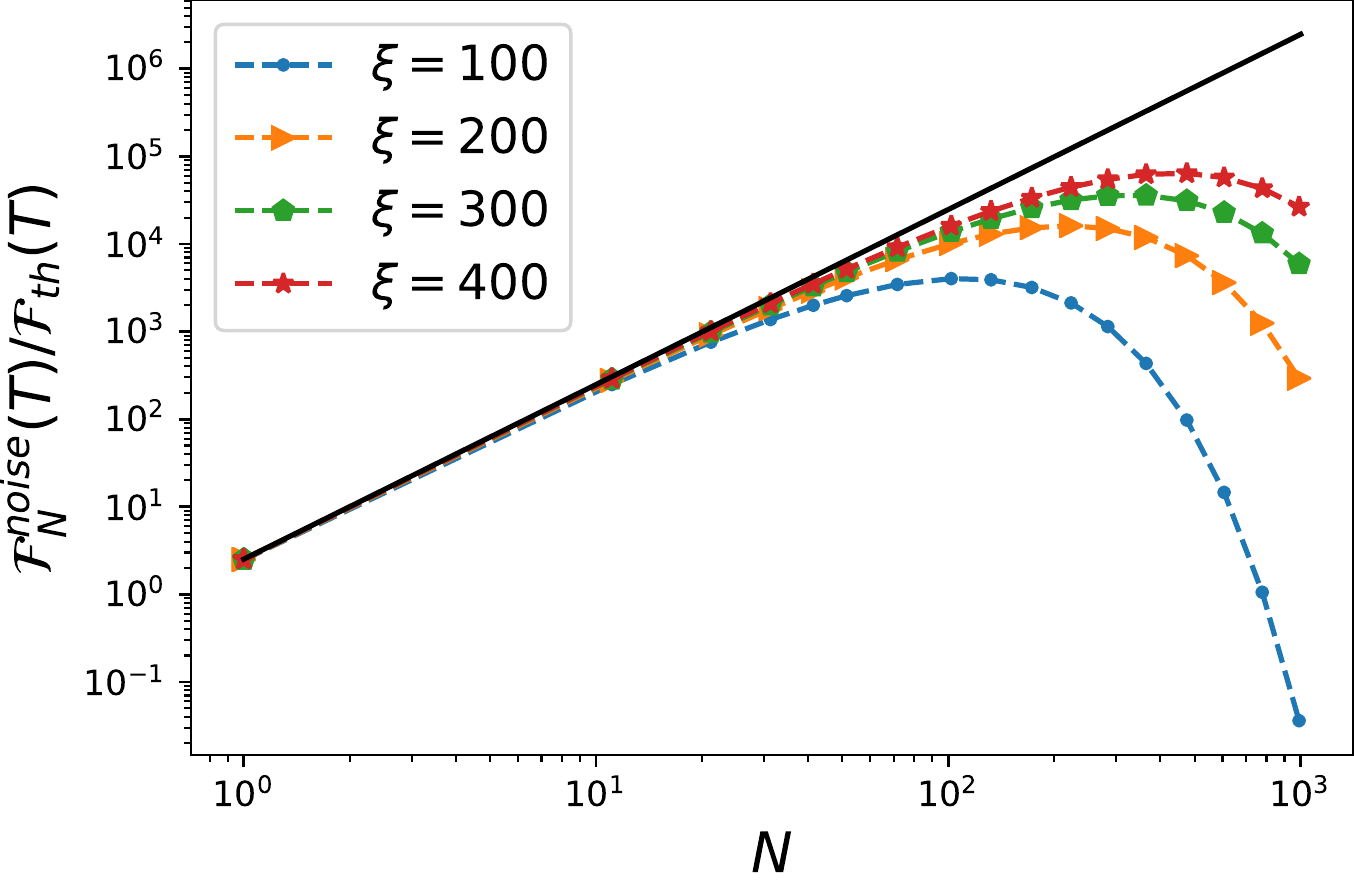}
	\caption{Log-log plot of the quantum Fisher information $\mathcal{F}_N^\textrm{noise}(T)$ as a function of $N$ in units of $\mathcal{F}_\textrm{th}(T)$ for different $\xi$. The black solid line corresponds to the quantum Fisher information in Eq.~(\ref{Example-QFI-ideal}). Parameters used are $k_B T/\hbar\Omega=2$ and $\eta=1/10$. }
	\label{fig2}
\end{figure}

The four dashed curves in Fig.~2 correspond to the QFIs associated with the four values of $\xi$ (see the figure legend).
As can be seen from this figure, these dashed curves approach the black solid line when $N$ is no larger than a certain $N_\textrm{max}$. This means that the HS described by Eq.~(\ref{Example-QFI-ideal}) is attained in the regime of $N\leq N_\textrm{max}$. It is easy to see that $N_\textrm{max}$ depends on $\xi$; more precisely, the larger $\xi$ is, the greater $N_\textrm{max}$ can be. We numerically find that $N_\textrm{max}$ may be respectively taken to be $109$, $217$, $326$, $435$ for the four values of $\xi$. Note that the HS achieved in current experiments is usually confined to the regime of $N\leq N_\textrm{max}=10$ \cite{Cimini} due to the detrimental effect of noises. The above numerical results demonstrate that our scheme allows for approaching HS in the presence of noises for a wide range of $N$.  Moreover, it is worth noting that strongly dissipative processes have been experimentally realized in a number of quantum platforms (see the review article \cite{2012Mueller1}). This means that the HS permitted in the above wide ranges of $N$ may be achieved using a variety
of experimental setups (to be listed in the Discussion section), which makes our scheme appealing from a practical perspective.

\begin{figure}
	\includegraphics[width=0.45\textwidth]{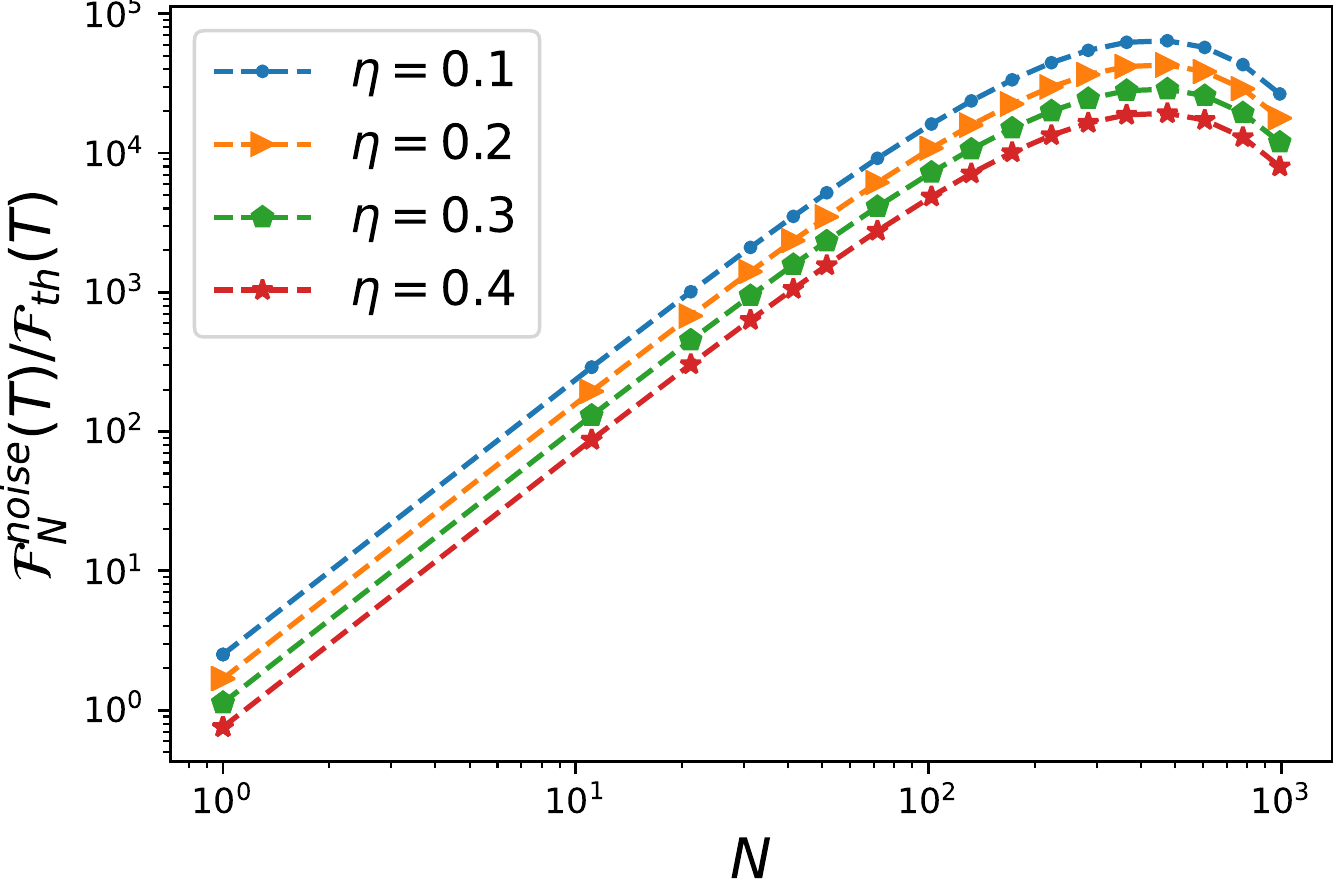}
	\caption{{Log-log plot of $\mathcal{F}_N^\textrm{noise}(T)$ as a function of $N$ in units of $\mathcal{F}_\textrm{th}(T)$ for different $\eta$. Parameters used are $k_B T/\hbar\Omega=2$ and $\xi=400$.} }
	\label{fig3}
\end{figure}

Figure 3 depicts $\mathcal{F}_N^\textrm{noise}(T)$ as a function of $N$ in units of $\mathcal{F}_\textrm{th}(T)$ for different noise strength $\eta$. Here, $k_B T/\hbar\Omega=2$, $\xi=400$, and four values of $\eta$ are chosen, namely, $\eta=0.1,0.2,0.3,0.4$. As can be seen from Fig.~3, $\mathcal{F}_N^\textrm{noise}(T)$ decreases as $\eta$ increases, which is expected from a physical point of view. It is interesting to note that $N_\textrm{max}$ does not change much in the course of varying $\eta$, indicating that the range in which the HS  is attained is irrespective of the value of $\eta$. Moreover, in Supplementary Note 3, we examine the influence of the initial state of the $N$ probes and the ancilla on $\mathcal{F}_N^\textrm{noise}(T)$. We consider the initial state of the form $\rho^{\otimes N}\otimes\sigma$, where $\rho$ and $\sigma$ are arbitrarily given $2\times 2$ density matrices. Both analytical and numerical results show that $\mathcal{F}_N^\textrm{noise}(T)$ does not depend much on $\rho$ but heavily relies on $\sigma$. More precisely, $\mathcal{F}_N^\textrm{noise}(T)$ grows quadratically  as a function of the $l_1$ norm of coherence of $\sigma$ \cite{2018Zhang170501}.

We remark that the HS given by Eq.~(\ref{Example-QFI-ideal}) is lost when $N$ is too large. The underlying reason is that there is some memory effect for each $\mathcal{E}_{\mathpzc{S}_n\mathpzc{A}}(t_1,t_0)$ appearing in decomposition (\ref{DP-decomposition}) of $\mathcal{E}_N(t_1,t_0)$ and these memory effects may add linearly  so that the total memory effect for $\mathcal{E}_N(t_1,t_0)$ becomes non-negligible for a too large $N$. Nevertheless, as long as the Markovian thermalization is strong enough, these memory effects can be made very weak and any large value of $N_\textrm{max}$ can be obtained in our scheme. By the way, considering that a stronger thermalization allows for producing a larger number of Gibbs states per unit time, a direct approach of taking advantage of strong thermalization for thermometry might be to repeatedly produce and measure Gibbs states with individual probes. Yet, this approach is ineffective in practice. Indeed, real-word quantum measurements suffer from correlated background noises if the measurement time is shorter than the noise correlation time \cite{2011Feizpour133603,2020Chu20501}. As these noises cannot be averaged out by repetitive measurements \cite{2011Feizpour133603,2020Chu20501}, the direct approach may not even be able to beat the standard quantum limit in the presence of these noises.

\section{Discussion}
The key to the implementation of our scheme is to realize the weak coupling of $N$ probes to a same ancilla. Apart from the magnetic resonance force microscopy, this kind of coupling is achievable in a variety
of experimental setups, including superconducting qubits coupled to a microwave resonator \cite{2013Hatridge178}, Rydberg atoms coupled to a microwave cavity \cite{2019Garcia193201}, trapped ions coupled to a common motional mode \cite{2015Pedernales15472} or an optical cavity mode \cite{2019Lee153603}, and nitrogen-vacancy (NV) centers in diamond coupled to a microwave mode in a superconducting transmission line cavity \cite{2017Astner140502}.

Thanks to the freedom in choosing explicit forms of $\mathcal{L}_\mathpzc{S}$ and $\rho_{_T}$, various kinds of probes can be used in our scheme. Particularly, the probes in previous proposals of thermometry may be used in our scheme, and our scheme may be exploited to promote these proposals to reach HS. To this end, one only needs to find an ancilla that is able to couple with the $N$ probes given in a previous proposal. The ancilla can be a NV center if hybridization proposals \cite{2018Wang11042} are under consideration, or a measuring apparatus when quantum nondemolition measurements are used \cite{2019Mehboudi30403,2020Pati12204}, or even an environment to which we have access \cite{2011Giovannetti222,2017Degen35002}.

The applications of our scheme are not limited to thermometry but can be straightforwardly extended to other metrological tasks. Indeed, our scheme is applicable to any metrological task in which the parameter of interest can be encoded in the unique steady state of a Markovian system. The metrological tasks of this kind are far more than quantum thermometry and are frequently encountered in quantum sensing, especially in noisy quantum metrology \cite{2011Giovannetti222}. Two examples are the detection of rotating fields \cite{Chen} and the sensing of low-frequency signals \cite{2020Xie14013}.

In summary, we have proposed a scheme of thermometry to approach HS for a wide range of $N$, based on the finding that an $N$-fold increase of the temperature-dependent phase can be obtained from the continuous interplay between the strong thermalization of $N$ probes and the weak coupling to the same external ancilla. As opposed to conventional ones, our scheme gets rid of a number of experimentally demanding requirements, e.g., the preparation of highly entangled states, and is implementable in a variety
of experimental setups. More importantly, it offers the key advantage of robustness irrespective of the type of noise in question, without resorting to complicated error correction techniques. Therefore, our scheme provides a feasible and robust pathway to the HS, with the entanglement-free feature which is {noteworthy in view of} previous proposals of Heisenberg-scalable thermometry. Two directions for future work are to exploit our scheme to promote previous proposals of thermometry to reach HS and to apply our scheme to other metrological tasks. Besides, throughout this paper, we have assumed the Markovian dissipative process described by the Lindblad equation, which is an approximative description relying on a number of simplifications. A more detailed analysis of our scheme starting from a fully microscopic description, while going beyond the scope of the present work, would be an interesting topic for future studies.

\section{Methods}
To obtain Eq.~(\ref{bound-memory-effect}), it is convenient to express $\mathcal{L}$ as
$\mathcal{L}=\mathcal{L}_\mathpzc{S}+g\mathcal{K}$ with $\mathcal{K}X:=-\textrm{i}[S\otimes A, X]$. Using the equalities $\mathcal{L}_\mathpzc{S}\mathcal{P}=\mathcal{P}\mathcal{L}_\mathpzc{S}=0$ \cite{2016Zhang12117,2016Zhang52132}, we have
\begin{eqnarray}\label{bound-step1}
\mathcal{PL}=g\mathcal{PK},~~~~\mathcal{LP}=g\mathcal{KP}.
\end{eqnarray}
Note that $\norm{\mathcal{E}_1\mathcal{E}_2}\leq\norm{\mathcal{E}_1}\norm{\mathcal{E}_2}$ for two superoperators $\mathcal{E}_1$ and $\mathcal{E}_2$. Here and throughout this paper, the Hilbert-Schmidt norm is adopted. That
is, for an operator $X$, the norm reads $\norm{X}:=\sqrt{\tr(X^\dagger X)}$,
and for a superoperator $\mathcal{E}$, it is the induced norm defined as
$\norm{\mathcal{E}}:=\sup_{\norm{X}\leq
1}\norm{\mathcal{E}(X)}$. It follows from Eq.~(\ref{bound-step1}) that $\norm{\mathcal{P}\mathcal{L}}\leq g\norm{\mathcal{P}}\norm{\mathcal{K}}$ and $\norm{\mathcal{L}\mathcal{P}}\leq g\norm{\mathcal{K}}\norm{\mathcal{P}}$. Using these two inequalities and noting that $\norm{\rho(s)}=\sqrt{\tr\rho^2(s)}\leq 1$, we have
\begin{eqnarray}\label{bound-step2}
&&\norm{\int_{t_0}^t ds\mathcal{P}\mathcal{L}
\mathcal{G}(t,s)\mathcal{Q}\mathcal{L}\mathcal{P}\rho(s)}
\leq \int_{t_0}^t ds\norm{\mathcal{P}\mathcal{L}
\mathcal{G}(t,s)\mathcal{Q}\mathcal{L}\mathcal{P}\rho(s)}\nonumber\\
&&\leq \norm{\mathcal{P}\mathcal{L}}\left(\int_{t_0}^tds \norm{\mathcal{G}(t,s)\mathcal{Q}}\right)
\norm{\mathcal{L}\mathcal{P}}
\leq\alpha g^2,
\end{eqnarray}
where $\alpha:=\norm{\mathcal{P}}^2\norm{\mathcal{K}}^2\int_{t_0}^tds \norm{\mathcal{G}(t,s)\mathcal{Q}}$. It can be shown that
\begin{eqnarray}\label{inequality-sm}
\norm{\mathcal{G}(t,s)\mathcal{Q}}\leq \varepsilon \exp[{\left(\varepsilon g\norm{\mathcal{K}}-\lambda\right)(t-s)}].
\end{eqnarray}
The proof of Eq.~(\ref{inequality-sm}) is very technical and presented in Supplementary Note 4, where the expression for $\varepsilon$ is given. Under the condition
\begin{eqnarray}
\lambda/g >{\varepsilon\norm{\mathcal{K}}},
\end{eqnarray}
$\norm{\mathcal{G}(t,s)\mathcal{Q}}$ decreases exponentially as $t-s$ increases. Then,
\begin{eqnarray}\label{g-tau}
\alpha\leq \norm{\mathcal{P}}^2\norm{\mathcal{K}}^2\left[\varepsilon/(\lambda-\varepsilon g\norm{\mathcal{K}})\right].
\end{eqnarray}
Inserting Eq.~(\ref{g-tau}) into Eq.~(\ref{bound-step2}), we can immediately obtain Eq.~(\ref{bound-memory-effect}).

\vspace{2em}
\noindent\textbf{\large Data availability}\\
Numerical data from the plots presented are available from D.-J.Z. upon
request.

\vspace{2em}
\noindent\textbf{\large Code availability}\\
The codes used for this study are available from D.-J.Z. upon
request.

%

\vspace{3em}
\noindent\textbf{\large Acknowledgements}\\
This work was supported by the National Natural Science Foundation of
China through Grant Nos.~11775129 and 12174224.

\vspace{2em}
\noindent\textbf{\large Author contributions}\\
All the authors contribute equally to this work.

\vspace{2em}
\noindent\textbf{\large Competing interests}\\
The authors declare no competing interests.

\vspace{2em}


\onecolumngrid
\clearpage

\renewcommand{\theequation}{\thesubsection S.\arabic{equation}}
\setcounter{equation}{0}

\def\OSQFIgeneralmain{{19}}

\def\Exampleoutputstate{{24}}

\def\inequalitysm{{32}}

\renewcommand{\figurename}{Supplementary Figure}

\begin{center}
\textbf{{\large SUPPLEMENTAL INFORMATION}}
\end{center}

\section*{Supplementary Note 1: Proof of Eq.~(\OSQFIgeneralmain) in the main text}

Here, we explain why the detrimental influence of noises on our scheme does not become increasingly severe as $N$ increases, irrespective of the type of noises in question.
By taking into account noises, the full dynamical equation of the $N$ probes and the ancilla reads
\begin{eqnarray}\label{eq:all}
	\partial_t\rho(t)=
	(\mathcal{L}_{\mathpzc{S}_1}+\mathcal{L}_{\mathpzc{S}_1}^{\textrm{noise}})\rho(t)
	-\textrm{i}g[S_1\otimes A, \rho(t)]+\cdots+
	(\mathcal{L}_{\mathpzc{S}_N}+\mathcal{L}_{\mathpzc{S}_N}^{\textrm{noise}})\rho(t)
	-\textrm{i}g[S_N\otimes A, \rho(t)]+\mathcal{L}_\mathpzc{A}^\textrm{noise}\rho(t):=\mathcal{L}\rho(t).\nonumber\\
\end{eqnarray}
Here, $\mathcal{L}_{\mathpzc{S}_n}^{\textrm{noise}}$ and $\mathcal{L}_{\mathpzc{A}}^{\textrm{noise}}$ denote the noises acting on the $n$th probe $\mathpzc{S}_n$ and the ancilla $\mathpzc{A}$, respectively. $S_n$ is a Hermitian operator acting on $\mathpzc{S}_n$. We assume for simplicity that $\mathcal{L}_{\mathpzc{S}_n}^{\textrm{noise}}$'s are identical with each other and $S_n$'s are identical with each other. Our following discussion can be straightforwardly extended to the scenario that $\mathcal{L}_{\mathpzc{S}_n}^{\textrm{noise}}$'s are different and $S_n$'s are different.

Let $\rho_{_T}^\prime$ be the steady state associated with the Liouvillian $\mathcal{L}_{\mathpzc{S}_n}+\mathcal{L}_{\mathpzc{S}_n}^{\textrm{noise}}$; that is,
$(\mathcal{L}_{\mathpzc{S}_n}+\mathcal{L}_{\mathpzc{S}_n}^{\textrm{noise}})\rho_{_T}^\prime=
0$. Define $\mathcal{P}$ as $\mathcal{P}X=\rho_{_T}^{\prime\otimes N}\otimes\tr_{\mathpzc{S}_1\cdots\mathpzc{S}_N}X$ and $\mathcal{Q}$ as $\mathcal{Q}X=X-\mathcal{P}X$. Using the same reasoning as in deriving Eq.~(5) in the main text, we have
\begin{eqnarray}\label{eq:relevant-closed-SM}
	\partial_t\mathcal{P}\rho(t)=\mathcal{P}\mathcal{L}\mathcal{P}\rho(t)+
	\int_{t_0}^t ds\mathcal{P}\mathcal{L}\mathcal{G}(t,s)\mathcal{Q}\mathcal{L}\mathcal{P}\rho(s),
\end{eqnarray}
with $\mathcal{L}$
given by Eq.~(\ref{eq:all}). As in the main text, the second term on the right-hand side of Eq.~(\ref{eq:relevant-closed-SM}) represents the total memory effect and is of the order $\mathcal{O}(g^2)$. Under the assumption that the Markovian thermalization of each probe ${\mathpzc{S}_n}$ is strong so that the total memory effect is negligible, we can neglect the second term on the right-hand side of Eq.~(\ref{eq:relevant-closed-SM}). Then, we have
\begin{eqnarray}\label{eq3-1}
	\partial_t\mathcal{P}\rho(t)=\mathcal{P}\mathcal{L}\mathcal{P}\rho(t).
\end{eqnarray}
Clearly,
\begin{eqnarray}\label{eq3-2}
	\mathcal{P}\rho(t)=\rho_{_T}^{\prime\otimes N}\otimes\rho_\mathpzc{A}(t).
\end{eqnarray}
A direct calculation shows that
\begin{eqnarray}\label{eq3-3}
	\mathcal{P}\mathcal{L}\mathcal{P}\rho(t)=\rho_{_T}^{\prime\otimes N}\otimes
	\left\{-\textrm{i}Ng[\varphi_{_T}^\prime A, \rho_\mathpzc{A}(t)]+\mathcal{L}_\mathpzc{A}^\textrm{noise}\rho_\mathpzc{A}(t)\right\},
\end{eqnarray}
where $\varphi_{_T}^\prime=\tr(S_i\rho_{_T}^\prime)$.
Substituting Eqs.~(\ref{eq3-2}) and (\ref{eq3-3}) into Eq.~(\ref{eq3-1}) yields
\begin{eqnarray}\label{eq3-6}
	\partial_t\rho_\mathpzc{A}(t)=-\textrm{i}Ng[\varphi_{_T}^\prime A, \rho_\mathpzc{A}(t)]+\mathcal{L}_\mathpzc{A}^\textrm{noise}\rho_\mathpzc{A}(t).
\end{eqnarray}
Note that $\mathcal{L}_{\mathpzc{S}_n}$ is assumed to be strong and, therefore, the steady state associated with $\mathcal{L}_{\mathpzc{S}_n}+\mathcal{L}_{\mathpzc{S}_n}^{\textrm{noise}}$ is mainly determined by $\mathcal{L}_{\mathpzc{S}_n}$. So,
\begin{eqnarray}\label{eq3-5}
	\rho_{_T}^\prime\approx\rho_{_T},
\end{eqnarray}
and hence,
\begin{eqnarray}
	\varphi_{_T}^\prime\approx\varphi_{_T}.
\end{eqnarray}
That is, the effect of $\mathcal{L}_{\mathpzc{S}_n}^{\textrm{noise}}$ is to slightly alter the steady state of $\mathpzc{S}_n$, which does not affect much the performance of our scheme.
Then, we can rewrite Eq.~(\ref{eq3-6}) as
\begin{eqnarray}\label{eq3-7}
	\partial_t\rho_\mathpzc{A}(t)=-\textrm{i}Ng[\varphi_{_T} A, \rho_\mathpzc{A}(t)]+\mathcal{L}_\mathpzc{A}^\textrm{noise}\rho_\mathpzc{A}(t).
\end{eqnarray}
$\mathcal{L}_\mathpzc{A}^\textrm{noise}$ on the right-hand side of Eq.~(\ref{eq3-7}) is independent of $N$. Also, no restriction is imposed on the form of $\mathcal{L}_\mathpzc{A}^\textrm{noise}$. These facts indicate that the detrimental influence of noises on our scheme does not become increasingly severe as $N$ increases, irrespective of the type of noises in question.

To confirm the above point, let us figure out the QFI associated with the output state of $\mathpzc{A}$. Without loss of generality, we can express $\mathcal{L}_\mathpzc{A}^\textrm{noise}$ as
\begin{eqnarray}\label{Lindblad-form}
	\mathcal{L}_\mathpzc{A}^\textrm{noise}\rho=\sum_l\kappa_l\left(K_l\rho K_l^\dagger
	-\frac{1}{2}\{K_l^\dagger K_l,\rho\}\right),
\end{eqnarray}
where $\kappa_l$ is the decaying rate and $K_l$ denotes the jump operator. Define
\begin{eqnarray}\label{rho-t-def}
	\widetilde{{\rho}}_\mathpzc{A}(t):=U^\dagger(t,t_0){\rho}_\mathpzc{A}(t)U(t,t_0),
\end{eqnarray}
where
$U(t,t_0)=e^{-\textrm{i}Ng\varphi_{_T}A(t-t_0)}$. Using Eqs.~(\ref{eq3-7}), (\ref{Lindblad-form}), and (\ref{rho-t-def}), we have that the dynamical equation for $\widetilde{\rho}_\mathpzc{A}(t)$ reads
\begin{eqnarray}\label{dynamical-t-rho}
	\partial_t\widetilde{\rho}_\mathpzc{A}(t)=\mathcal{L}_\mathpzc{A}^\textrm{noise}(t,t_0)
	\widetilde{\rho}_\mathpzc{A}(t),
\end{eqnarray}
where
\begin{eqnarray}\label{L-t}
	\mathcal{L}_\mathpzc{A}^\textrm{noise}(t,t_0)\rho=
	\sum_l\kappa_l\left[{K}_l(t,t_0) \rho {K}_l^\dagger(t,t_0)
	-\frac{1}{2}\left\{{K}_l^\dagger(t,t_0) {K}_l(t,t_0),\rho\right\}\right],
\end{eqnarray}
with
\begin{eqnarray}
	{K}_l(t,t_0)=U^\dagger(t,t_0)K_lU(t,t_0).
\end{eqnarray}
Noting that
\begin{eqnarray}
	U(t,t_0)=\sum_\mu e^{-\textrm{i}Ng\varphi_{_T}a_\mu(t-t_0)}\ket{a_\mu}\bra{a_\mu},
\end{eqnarray}
where $a_\mu$ and $\ket{a_\mu}$ denote the eigenvalue and the associated eigenstate of $A$, we have that
\begin{eqnarray}\label{t-K}
	{K}_l(t,t_0)=\sum_{\mu,\nu}e^{\textrm{i}Ng\varphi_{_T}(a_\mu-a_\nu)(t-t_0)}
	K_{l\mu\nu}\ket{a_\mu}\bra{a_\nu},
\end{eqnarray}
with
\begin{eqnarray}
	K_{l\mu\nu}=\bra{a_\mu}K_l\ket{a_\nu}
\end{eqnarray}
Substituting Eq.~(\ref{t-K}) into Eq.~(\ref{L-t}) yields
\begin{eqnarray}\label{L-t-2}
	&&\mathcal{L}_\mathpzc{A}^\textrm{noise}(t,t_0)\rho=\nonumber\\
	&&\sum_l\kappa_l\sum_{\mu,\nu,\mu^\prime,\nu^\prime}
	e^{\textrm{i}Ng\varphi_{_T}(a_\mu-a_\nu-
		a_\mu^\prime+a_\nu^\prime)(t-t_0)}
	K_{l\mu\nu}K_{l\mu^\prime\nu^\prime}^*
	\left(\ket{a_\mu}\bra{a_\nu}\rho\ket{a_\nu^\prime}\bra{a_\mu^\prime}
	-\frac{1}{2}\left\{\ket{a_\nu^\prime}\bra{a_\mu^\prime}\ket{a_\mu}\bra{a_\nu}, \rho\right\}\right).
\end{eqnarray}
Note that, for a large $N$, the complex exponentials $e^{\textrm{i}Ng\varphi_{_T}(a_\mu-a_\nu-
	a_\mu^\prime+a_\nu^\prime)(t-t_0)}$ are rapidly oscillating when $a_\mu-a_\nu\neq a_\mu^\prime-a_\nu^\prime$. Hence, on an appreciable time scale, the oscillations will quickly average to zero. Using the rotating wave approximation, we can neglect the terms in Eq.~(\ref{L-t-2}) for $a_\mu-a_\nu\neq a_\mu^\prime-a_\nu^\prime$ and, therefore, obtain the following effective Liouvillian,
\begin{eqnarray}\label{L-t-3}
	\widetilde{\mathcal{L}}_{\mathpzc{A}}^\textrm{noise}\rho=
	\sum_l\kappa_l\sum_{a_\mu-a_\nu= a_\mu^\prime-a_\nu^\prime}
	K_{l\mu\nu}K_{l\mu^\prime\nu^\prime}^*
	\left(\ket{a_\mu}\bra{a_\nu}\rho\ket{a_\nu^\prime}\bra{a_\mu^\prime}
	-\frac{1}{2}\left\{\ket{a_\nu^\prime}\bra{a_\mu^\prime}\ket{a_\mu}\bra{a_\nu}, \rho\right\}\right).
\end{eqnarray}
Let $I:=\{a_\mu-a_\nu| \textrm{$a_\mu$ and $a_\nu$ are the eigenvalues of $A$}\}$ be the set comprised of the differences among the eigenvalues of $A$. We can rewrite Eq.~(\ref{L-t-3}) as
\begin{eqnarray}\label{L-t-eff}
	\widetilde{\mathcal{L}}_{\mathpzc{A}}^\textrm{noise}\rho=
	\sum_l\sum_{\Delta\in I}\kappa_l\left(K_{l,\Delta}\rho K_{l,\Delta}^\dagger
	-\frac{1}{2}\left\{K_{l,\Delta}^\dagger K_{l,\Delta},\rho\right\}\right),
\end{eqnarray}
with
\begin{eqnarray}
	K_{l,\Delta}:=\sum_{a_\mu-a_\nu=\Delta}K_{l\mu\nu}\ket{a_\mu}\bra{a_\nu}=
	\sum_{a_\mu-a_\nu=\Delta}\ket{a_\mu}\bra{a_\mu}K_l\ket{a_\nu}\bra{a_\nu}.
\end{eqnarray}
Hence,
\begin{eqnarray}
	\widetilde{\rho}_\mathpzc{A}(t)=
	e^{\widetilde{\mathcal{L}}_{\mathpzc{A}}^\textrm{noise}(t-t_0)}
	\widetilde{\rho}_\mathpzc{A}(t_0).
\end{eqnarray}
Noting that $\rho_\mathpzc{A}(t)=U(t,t_0)\widetilde{\rho}_\mathpzc{A}(t)U^\dagger(t,t_0)$ and $\widetilde{\rho}_\mathpzc{A}(t_0)={\rho_\mathpzc{A}}(t_0)=\frac{1}{2}(\ket{a_M}+\ket{a_m})
(\bra{a_M}+\bra{a_m})$, we have that the output state of the ancilla reads
\begin{eqnarray}
	\rho_\mathpzc{A}(t_0+1/g)=U^{-\textrm{i}N\varphi_{_T}A}\delta U^{\textrm{i}N\varphi_{_T}A}.
\end{eqnarray}
Here $\delta$ is defined to be the state
\begin{eqnarray}
	\delta=e^{\widetilde{\mathcal{L}}_{\mathpzc{A}}^\textrm{noise}/g}\left
	[\frac{1}{2}(\ket{a_M}+\ket{a_m})
	(\bra{a_M}+\bra{a_m})\right],
\end{eqnarray}
whose eigendecomposition can be expressed as
\begin{eqnarray}
	\delta=\sum_kp_k\ket{\phi_k}\bra{\phi_k}.
\end{eqnarray}
Then, the QFI associated with the output state reads
\begin{eqnarray}\label{OS-QFI-general}
	\mathcal{F}_N^\textrm{noise}(T)=\left(2\sum_{k\neq l}p_{k,l}\abs{\bra{\phi_k}A\ket{\phi_l}}^2\right)
	N^2 \abs{\partial_{_T}\varphi_{_T}}^2,
\end{eqnarray}
with coefficients
\begin{eqnarray}
	p_{k,l}=
	\begin{cases}
		0, & \mbox{if } p_k=p_l=0 \\
		\frac{(p_k-p_l)^2}{p_k+p_l}, & \mbox{otherwise}.
	\end{cases}
\end{eqnarray}

\section*{Supplementary Note 2: Proof of Eq.~(\Exampleoutputstate) in the main text}

Here, we figure out the expression of $\rho_\mathpzc{A}(t_1)$ for the model under consideration. Upon taking into account the dephasing noise, the full dynamics of the $N$ probes and the ancilla reads
\begin{eqnarray}
	\partial_t\rho(t)=\left[(\mathcal{L}_{\mathpzc{S}_1\mathpzc{A}}+
	\mathcal{L}_{\mathpzc{S}_1}^\textrm{noise})
	+\cdots+(\mathcal{L}_{\mathpzc{S}_N\mathpzc{A}}+
	\mathcal{L}_{\mathpzc{S}_N}^\textrm{noise})
	+\mathcal{L}_\mathpzc{A}^\textrm{noise}\right]\rho(t),
\end{eqnarray}
where the expressions for $\mathcal{L}_{\mathpzc{S}_n\mathpzc{A}}$, $\mathcal{L}_{\mathpzc{S}_n}^\textrm{noise}$, and $\mathcal{L}_\mathpzc{A}^\textrm{noise}$ are given in the main text.
It is worth noting that the Liouville superoperators $\mathcal{L}_{\mathpzc{S}_n\mathpzc{A}}+
\mathcal{L}_{\mathpzc{S}_n}^\textrm{noise}$ commute with each other and, moreover, they commute with the Liouville superoperator $\mathcal{L}_\mathpzc{A}^\textrm{noise}$,
\begin{eqnarray}
	[\mathcal{L}_{\mathpzc{S}_n\mathpzc{A}}+
	\mathcal{L}_{\mathpzc{S}_n}^\textrm{noise}, \mathcal{L}_{\mathpzc{S}_m\mathpzc{A}}+
	\mathcal{L}_{\mathpzc{S}_m}^\textrm{noise}]=0,~~~~
	[\mathcal{L}_{\mathpzc{S}_n\mathpzc{A}}+
	\mathcal{L}_{\mathpzc{S}_n}^\textrm{noise},
	\mathcal{L}_{\mathpzc{A}}^\textrm{noise}]=0.
\end{eqnarray}
Hence, the dynamical map $\mathcal{E}_N(t_1,t_0)=\exp\left[\left(\mathcal{L}_{\mathpzc{S}_1\mathpzc{A}}+
\mathcal{L}_{\mathpzc{S}_1}^\textrm{noise}
+\cdots+\mathcal{L}_{\mathpzc{S}_N\mathpzc{A}}+
\mathcal{L}_{\mathpzc{S}_N}^\textrm{noise}
+\mathcal{L}_\mathpzc{A}^\textrm{noise}\right)(t_1-t_0)\right]$ can be decomposed as
\begin{eqnarray}\label{DP-decomposition-SM}
	\mathcal{E}_N(t_1,t_0)=\mathcal{E}_{\mathpzc{S}_1\mathpzc{A}}(t_1,t_0)\cdots
	\mathcal{E}_{\mathpzc{S}_N\mathpzc{A}}(t_1,t_0)\mathcal{E}_{\mathpzc{A}}(t_1,t_0),
\end{eqnarray}
where
\begin{eqnarray}
	\mathcal{E}_{\mathpzc{S}_n\mathpzc{A}}(t_1,t_0)=\exp[
	\left(\mathcal{L}_{\mathpzc{S}_n\mathpzc{A}}+\mathcal{L}_{\mathpzc{S}_n}^\textrm{noise}\right)
	(t_1-t_0)],~~~~
	\mathcal{E}_{\mathpzc{A}}(t_1,t_0)=\exp[\mathcal{L}_\mathpzc{A}^\textrm{noise}(t_1-t_0)],
\end{eqnarray}
with $t_1=t_0+1/g$.
Noting that $\rho_{\mathpzc{A}}(t_0)=(\ket{0}+\ket{1})
(\bra{0}+\bra{1})/2$ and $\rho_\mathpzc{A}(t_1)=\tr_{\mathpzc{S}_1\cdots\mathpzc{S}_N}
\left[\mathcal{E}_N(t_1,t_0)\rho_{_T}^{\otimes N}\otimes\rho_\mathpzc{A}(t_0)\right]$, we have
\begin{eqnarray}\label{note2-four-terms}
	\rho_\mathpzc{A}(t_1)&=&\frac{1}{2}\tr_{\mathpzc{S}_1\cdots\mathpzc{S}_N}
	\left[\mathcal{E}_N(t_1,t_0)\rho_{_T}^{\otimes N}\otimes\ket{0}\bra{0}\right]
	+
	\frac{1}{2}\tr_{\mathpzc{S}_1\cdots\mathpzc{S}_N}
	\left[\mathcal{E}_N(t_1,t_0)\rho_{_T}^{\otimes N}\otimes\ket{1}\bra{1}\right]\nonumber\\
	&+&
	\frac{1}{2}\tr_{\mathpzc{S}_1\cdots\mathpzc{S}_N}
	\left[\mathcal{E}_N(t_1,t_0)\rho_{_T}^{\otimes N}\otimes\ket{0}\bra{1}\right]
	+
	\frac{1}{2}\tr_{\mathpzc{S}_1\cdots\mathpzc{S}_N}
	\left[\mathcal{E}_N(t_1,t_0)\rho_{_T}^{\otimes N}\otimes\ket{1}\bra{0}\right].
\end{eqnarray}
In the following, we make use of the decomposition in Eq.~(\ref{DP-decomposition-SM}) to figure out all of the four terms on the right-hand side of Eq.~(\ref{note2-four-terms}).

{First, we figure out the term $\frac{1}{2}\tr_{\mathpzc{S}_1\cdots\mathpzc{S}_N}
	\left[\mathcal{E}_N(t_1,t_0)\rho_{_T}^{\otimes N}\otimes\ket{0}\bra{0}\right]$.} It is easy to see that $\mathcal{E}_{\mathpzc{A}}(t_1,t_0)$ maps the state $\rho_{_T}^{\otimes N}\otimes\ket{0}\bra{0}$ into itself,
\begin{eqnarray}
	\rho_{_T}^{\otimes N}\otimes\ket{0}\bra{0}\xrightarrow{\mathcal{E}_{\mathpzc{A}}(t_1,t_0)}
	\rho_{_T}^{\otimes N}\otimes\ket{0}\bra{0}.
\end{eqnarray}
Hence,
\begin{eqnarray}\label{nt2-s1}
	\mathcal{E}_N(t_1,t_0)\rho_{_T}^{\otimes N}\otimes\ket{0}\bra{0}=\mathcal{E}_{\mathpzc{S}_1\mathpzc{A}}(t_1,t_0)\cdots
	\mathcal{E}_{\mathpzc{S}_N\mathpzc{A}}(t_1,t_0)
	\rho_{_T}^{\otimes N}\otimes\ket{0}\bra{0}.
\end{eqnarray}
Noting that
\begin{eqnarray}
	\left(\mathcal{L}_{\mathpzc{S}_n\mathpzc{A}}+
	\mathcal{L}_{\mathpzc{S}_n}^\textrm{noise}\right)\rho\otimes\ket{0}\bra{0}
	=\left(\mathcal{L}_{\mathpzc{S}_n}\rho
	-\textrm{i}g[\sigma_z^{\mathpzc{S}_n},\rho]
	+\mathcal{L}_{\mathpzc{S}_n}^\textrm{noise}\rho\right)\otimes\ket{0}\bra{0}
	=:\mathcal{L}_{\mathpzc{S}_n\mathpzc{A}}^{(00)}\rho\otimes\ket{0}\bra{0},
\end{eqnarray}
where $\mathcal{L}_{\mathpzc{S}_n\mathpzc{A}}^{(00)}\rho=\mathcal{L}_{\mathpzc{S}_n}\rho
-\textrm{i}g[\sigma_z^{\mathpzc{S}_n},\rho]
+\mathcal{L}_{\mathpzc{S}_n}^\textrm{noise}\rho$ is a Liouville superoperator acting on $\mathpzc{S}_n$, we have
\begin{eqnarray}
	\mathcal{E}_{\mathpzc{S}_n\mathpzc{A}}(t_1,t_0)\rho_{_T}\otimes\ket{0}\bra{0}
	=\exp[\mathcal{L}_{\mathpzc{S}_n\mathpzc{A}}^{(00)}(t_1-t_0)]\rho_{_T}\otimes\ket{0}\bra{0}.
\end{eqnarray}
Since $\mathcal{L}_{\mathpzc{S}_n\mathpzc{A}}^{(00)}$ is in the Lindblad form, $\exp[\mathcal{L}_{\mathpzc{S}_n\mathpzc{A}}^{(00)}(t_1-t_0)]$ is a completely positive and trace-preserving (CPTP) map. Then, we have
\begin{eqnarray}\label{nt2-s2}
	\tr_{\mathpzc{S}_n}
	\left[\mathcal{E}_{\mathpzc{S}_n\mathpzc{A}}(t_1,t_0)\rho_{_T}\otimes\ket{0}\bra{0}\right]
	=\tr_{\mathpzc{S}_n}
	\left[e^{\mathcal{L}_{\mathpzc{S}_n\mathpzc{A}}^{(00)}(t_1-t_0)}\rho_{_T}\right]\ket{0}\bra{0}
	=\ket{0}\bra{0}.
\end{eqnarray}
Using Eqs.~(\ref{nt2-s1}) and (\ref{nt2-s2}), we have
\begin{eqnarray}\label{nt2-term1}
	\frac{1}{2}\tr_{\mathpzc{S}_1\cdots\mathpzc{S}_N}
	\left[\mathcal{E}_N(t_1,t_0)\rho_{_T}^{\otimes N}\otimes\ket{0}\bra{0}\right]=\frac{1}{2}
	\ket{0}\bra{0}.
\end{eqnarray}

{Second, we figure out the term $\frac{1}{2}\tr_{\mathpzc{S}_1\cdots\mathpzc{S}_N}
	\left[\mathcal{E}_N(t_1,t_0)\rho_{_T}^{\otimes N}\otimes\ket{1}\bra{1}\right]$.} Likewise, $\mathcal{E}_{\mathpzc{A}}(t_1,t_0)$ maps the state $\rho_{_T}^{\otimes N}\otimes\ket{1}\bra{1}$ into itself,
\begin{eqnarray}
	\rho_{_T}^{\otimes N}\otimes\ket{1}\bra{1}\xrightarrow{\mathcal{E}_{\mathpzc{A}}(t_1,t_0)}
	\rho_{_T}^{\otimes N}\otimes\ket{1}\bra{1}.
\end{eqnarray}
Hence,
\begin{eqnarray}\label{nt2-s3}
	\mathcal{E}_N(t_1,t_0)\rho_{_T}^{\otimes N}\otimes\ket{1}\bra{1}=\mathcal{E}_{\mathpzc{S}_1\mathpzc{A}}(t_1,t_0)\cdots
	\mathcal{E}_{\mathpzc{S}_N\mathpzc{A}}(t_1,t_0)
	\rho_{_T}^{\otimes N}\otimes\ket{1}\bra{1}.
\end{eqnarray}
Noting that
\begin{eqnarray}
	\left(\mathcal{L}_{\mathpzc{S}_n\mathpzc{A}}+
	\mathcal{L}_{\mathpzc{S}_n}^\textrm{noise}\right)\rho\otimes\ket{1}\bra{1}
	=\left(\mathcal{L}_{\mathpzc{S}_n}\rho
	+\textrm{i}g[\sigma_z^{\mathpzc{S}_n},\rho]
	+\mathcal{L}_{\mathpzc{S}_n}^\textrm{noise}\rho\right)\otimes\ket{1}\bra{1}
	=:\mathcal{L}_{\mathpzc{S}_n\mathpzc{A}}^{(11)}\rho\otimes\ket{1}\bra{1},
\end{eqnarray}
where $\mathcal{L}_{\mathpzc{S}_n\mathpzc{A}}^{(11)}\rho=\mathcal{L}_{\mathpzc{S}_n}\rho
+\textrm{i}g[\sigma_z^{\mathpzc{S}_n},\rho]
+\mathcal{L}_{\mathpzc{S}_n}^\textrm{noise}\rho$, we have
\begin{eqnarray}
	\mathcal{E}_{\mathpzc{S}_n\mathpzc{A}}(t_1,t_0)\rho_{_T}\otimes\ket{1}\bra{1}
	=\exp[\mathcal{L}_{\mathpzc{S}_n\mathpzc{A}}^{(11)}(t_1-t_0)]\rho_{_T}\otimes\ket{1}\bra{1}.
\end{eqnarray}
Noting that $\mathcal{L}_{\mathpzc{S}_n\mathpzc{A}}^{(11)}$ is in the Lindblad form and $\exp[\mathcal{L}_{\mathpzc{S}_n\mathpzc{A}}^{(11)}(t_1-t_0)]$ is a CPTP map, we have
\begin{eqnarray}\label{nt2-s4}
	\tr_{\mathpzc{S}_n}
	\left[\mathcal{E}_{\mathpzc{S}_n\mathpzc{A}}(t_1,t_0)\rho_{_T}\otimes\ket{1}\bra{1}\right]
	=\tr_{\mathpzc{S}_n}
	\left[e^{\mathcal{L}_{\mathpzc{S}_n\mathpzc{A}}^{(11)}(t_1-t_0)}\rho_{_T}\right]\ket{1}\bra{1}
	=\ket{1}\bra{1}.
\end{eqnarray}
Using Eqs.~(\ref{nt2-s3}) and (\ref{nt2-s4}), we have
\begin{eqnarray}\label{nt2-term2}
	\frac{1}{2}\tr_{\mathpzc{S}_1\cdots\mathpzc{S}_N}
	\left[\mathcal{E}_N(t_1,t_0)\rho_{_T}^{\otimes N}\otimes\ket{1}\bra{1}\right]=\frac{1}{2}
	\ket{1}\bra{1}.
\end{eqnarray}

{Third, we figure out the term $\frac{1}{2}\tr_{\mathpzc{S}_1\cdots\mathpzc{S}_N}
	\left[\mathcal{E}_N(t_1,t_0)\rho_{_T}^{\otimes N}\otimes\ket{0}\bra{1}\right]$.} It is easy to see that
\begin{eqnarray}
	\rho_{_T}^{\otimes N}\otimes\ket{0}\bra{1}\xrightarrow{\mathcal{E}_{\mathpzc{A}}(t_1,t_0)}
	e^{-2\eta}\rho_{_T}^{\otimes N}\otimes\ket{0}\bra{1},
\end{eqnarray}
where $\eta=\kappa_\mathpzc{A}/g$.
Hence,
\begin{eqnarray}\label{nt2-s5}
	\mathcal{E}_N(t_1,t_0)\rho_{_T}^{\otimes N}\otimes\ket{0}\bra{1}=e^{-2\eta}\mathcal{E}_{\mathpzc{S}_1\mathpzc{A}}(t_1,t_0)\cdots
	\mathcal{E}_{\mathpzc{S}_N\mathpzc{A}}(t_1,t_0)
	\rho_{_T}^{\otimes N}\otimes\ket{0}\bra{1}.
\end{eqnarray}
Noting that
\begin{eqnarray}
	\left(\mathcal{L}_{\mathpzc{S}_n\mathpzc{A}}+
	\mathcal{L}_{\mathpzc{S}_n}^\textrm{noise}\right)\rho\otimes\ket{0}\bra{1}
	=\left(\mathcal{L}_{\mathpzc{S}_n}\rho
	-\textrm{i}g\{\sigma_z^{\mathpzc{S}_n},\rho\}
	+\mathcal{L}_{\mathpzc{S}_n}^\textrm{noise}\rho\right)\otimes\ket{0}\bra{1}
	=:\mathcal{L}_{\mathpzc{S}_n\mathpzc{A}}^{(01)}\rho\otimes\ket{0}\bra{1},
\end{eqnarray}
where $\mathcal{L}_{\mathpzc{S}_n\mathpzc{A}}^{(01)}\rho=\mathcal{L}_{\mathpzc{S}_n}\rho
-\textrm{i}g\{\sigma_z^{\mathpzc{S}_n},\rho\}
+\mathcal{L}_{\mathpzc{S}_n}^\textrm{noise}\rho$, we have
\begin{eqnarray}
	\mathcal{E}_{\mathpzc{S}_n\mathpzc{A}}(t_1,t_0)\rho_{_T}\otimes\ket{0}\bra{1}
	=\exp[\mathcal{L}_{\mathpzc{S}_n\mathpzc{A}}^{(01)}(t_1-t_0)]\rho_{_T}\otimes\ket{0}\bra{1}.
\end{eqnarray}
Then, we have
\begin{eqnarray}\label{nt2-s6}
	\tr_{\mathpzc{S}_n}
	\left[\mathcal{E}_{\mathpzc{S}_n\mathpzc{A}}(t_1,t_0)\rho_{_T}\otimes\ket{0}\bra{1}\right]
	=\tr_{\mathpzc{S}_n}
	\left[e^{\mathcal{L}_{\mathpzc{S}_n\mathpzc{A}}^{(01)}(t_1-t_0)}\rho_{_T}\right]\ket{0}\bra{1}
	=\Gamma\ket{0}\bra{1},
\end{eqnarray}
where $\Gamma=\tr_{\mathpzc{S}_n}
\left[e^{\mathcal{L}_{\mathpzc{S}_n\mathpzc{A}}^{(01)}(t_1-t_0)}\rho_{_T}\right]$. Using \textit{MATHEMATICA}, we can work out the expression for $\Gamma$, given by
\begin{eqnarray}\label{Example-Gamma}
	\Gamma=\frac{e^{\omega_+(\xi)}}{2}
	\left[1+\frac{(2\overline{n}+1)\xi}{\sqrt{\Delta(\xi)}}-
	\frac{4\textrm{i}}{(2\overline{n}+1)\sqrt{\Delta(\xi)}}\right]
	+\frac{e^{\omega_-(\xi)}}{2}
	\left[1-\frac{(2\overline{n}+1)\xi}{\sqrt{\Delta(\xi)}}+
	\frac{4\textrm{i}}{(2\overline{n}+1)\sqrt{\Delta(\xi)}}\right],
\end{eqnarray}
where 
\begin{eqnarray}\label{Delta}
	\Delta(\xi)=(2\overline{n}+1)^2\xi^2-8\textrm{i}\xi-16,
\end{eqnarray}
and 
\begin{eqnarray}\label{omega}
	\omega_\pm(\xi)=[-(2\overline{n}+1)\xi\pm\sqrt{\Delta(\xi)}]/2
\end{eqnarray}
with $\xi=\gamma/g$.
Using Eqs.~(\ref{nt2-s5}) and (\ref{nt2-s6}), we have
\begin{eqnarray}\label{nt2-term3}
	\frac{1}{2}\tr_{\mathpzc{S}_1\cdots\mathpzc{S}_N}
	\left[\mathcal{E}_N(t_1,t_0)\rho_{_T}^{\otimes N}\otimes\ket{0}\bra{1}\right]=\frac{1}{2}\Gamma^N e^{-2\eta}
	\ket{0}\bra{1}.
\end{eqnarray}

Now, substituting Eqs.~(\ref{nt2-term1}), (\ref{nt2-term2}), and (\ref{nt2-term3}) into Eq.~(\ref{note2-four-terms}), we have
\begin{eqnarray}
	\rho_\mathpzc{A}(t_1)=
	\frac{1}{2}
	\begin{pmatrix}
		1 & \Gamma^N e^{-2\eta} \\
		\Gamma^{*N} e^{-2\eta} & 1
	\end{pmatrix},
\end{eqnarray}
where we have used the fact that the term $\frac{1}{2}\tr_{\mathpzc{S}_1\cdots\mathpzc{S}_N}
\left[\mathcal{E}_N(t_1,t_0)\rho_{_T}^{\otimes N}\otimes\ket{0}\bra{1}\right]$ is complex conjugate to the term $\frac{1}{2}\tr_{\mathpzc{S}_1\cdots\mathpzc{S}_N}
\left[\mathcal{E}_N(t_1,t_0)\rho_{_T}^{\otimes N}\otimes\ket{1}\bra{0}\right]$. This completes the proof of Eq.~(\Exampleoutputstate).

\section*{Supplementary Note 3: Influence of the initial state on the quantum Fisher information}

Here, focusing on the model in the main text, we examine the influence of the initial state of the $N$ probes and the ancilla on the  quantum Fisher information (QFI) for the output state of the ancilla $\rho_\mathpzc{A}(t_1)$. We consider the initial state of the form
\begin{eqnarray}\label{Example-initial-state}
	\rho^{\otimes N}\otimes\sigma,
\end{eqnarray} 
where $\rho$ and $\sigma$ are arbitrarily given $2\times 2$ density matrices. In the following, we  figure out the analytical expression of the QFI for the output state of the ancilla and then examine the influence of the initial state given by Eq.~(\ref{Example-initial-state}) on the QFI.

Writing $\sigma$ as $\sigma=\sigma_{00}\ket{0}\bra{0}+\sigma_{11}\ket{1}\bra{1}+\sigma_{01}\ket{0}\bra{1}+\sigma_{01}^*\ket{1}\bra{0}$, we have 
\begin{eqnarray}\label{nt3-rho1}
	\rho_\mathpzc{A}(t_1)&=&\sigma_{00}\tr_{\mathpzc{S}_1\cdots\mathpzc{S}_N}
	\left[\mathcal{E}_N(t_1,t_0)\rho^{\otimes N}\otimes\ket{0}\bra{0}\right]
	+
	\sigma_{11}\tr_{\mathpzc{S}_1\cdots\mathpzc{S}_N}
	\left[\mathcal{E}_N(t_1,t_0)\rho^{\otimes N}\otimes\ket{1}\bra{1}\right]\nonumber\\
	&+&
	\sigma_{01}\tr_{\mathpzc{S}_1\cdots\mathpzc{S}_N}
	\left[\mathcal{E}_N(t_1,t_0)\rho^{\otimes N}\otimes\ket{0}\bra{1}\right]
	+
	\sigma_{01}^*\tr_{\mathpzc{S}_1\cdots\mathpzc{S}_N}
	\left[\mathcal{E}_N(t_1,t_0)\rho^{\otimes N}\otimes\ket{1}\bra{0}\right].
\end{eqnarray}
Using the same reasoning as in deriving Eqs.~(\ref{nt2-term1}), (\ref{nt2-term2}), and (\ref{nt2-term3}), we have 
\begin{eqnarray}
	\tr_{\mathpzc{S}_1\cdots\mathpzc{S}_N}
	\left[\mathcal{E}_N(t_1,t_0)\rho^{\otimes N}\otimes\ket{0}\bra{0}\right]=1,
	\label{nt3-term1}\\
	\tr_{\mathpzc{S}_1\cdots\mathpzc{S}_N}
	\left[\mathcal{E}_N(t_1,t_0)\rho^{\otimes N}\otimes\ket{1}\bra{1}\right]=1,
	\label{nt3-term2}\\
	\tr_{\mathpzc{S}_1\cdots\mathpzc{S}_N}
	\left[\mathcal{E}_N(t_1,t_0)\rho^{\otimes N}\otimes\ket{0}\bra{1}\right]=\Gamma^N e^{-2\eta}, \label{nt3-term3}
\end{eqnarray}
where
\begin{eqnarray}
	\Gamma=\tr_{\mathpzc{S}_n}
	\left[e^{\mathcal{L}_{\mathpzc{S}_n\mathpzc{A}}^{(01)}(t_1-t_0)}\rho\right],
\end{eqnarray}
with  $\mathcal{L}_{\mathpzc{S}_n\mathpzc{A}}^{(01)}\rho=\mathcal{L}_{\mathpzc{S}_n}\rho
-\textrm{i}g\{\sigma_z^{\mathpzc{S}_n},\rho\}
+\mathcal{L}_{\mathpzc{S}_n}^\textrm{noise}\rho$. Using \textit{MATHEMATICA}, we can work out the expression for $\Gamma$, given by
\begin{eqnarray}\label{nt3-Gamma}
	\Gamma=\frac{e^{\omega_+(\xi)}}{2}
	\left[1+\frac{(2\overline{n}+1)\xi}{\sqrt{\Delta(\xi)}}-
	\frac{4\textrm{i}}{\sqrt{\Delta(\xi)}}(\rho_{00}-\rho_{11})\right]
	+\frac{e^{\omega_-(\xi)}}{2}
	\left[1-\frac{(2\overline{n}+1)\xi}{\sqrt{\Delta(\xi)}}+
	\frac{4\textrm{i}}{\sqrt{\Delta(\xi)}}(\rho_{00}-\rho_{11})\right],
\end{eqnarray}
where we have expressed $\rho$ as $\rho=\rho_{00}\ket{0}\bra{0}+\rho_{11}\ket{1}\bra{1}+\rho_{01}\ket{0}\bra{1}+\rho_{01}^*\ket{1}\bra{0}$ and $\Delta(\xi)$ and $\omega_\pm(\xi)$ are given by Eqs.~(\ref{Delta}) and (\ref{omega}), respectively.
Substituting Eqs.~(\ref{nt3-term1}), (\ref{nt3-term2}), and (\ref{nt3-term3}) into 
Eq.~(\ref{nt3-rho1}), we have the output state of the ancilla 
\begin{eqnarray}
	\rho_\mathpzc{A}(t_1)=
	\begin{pmatrix}
		\sigma_{00} & \sigma_{01}\Gamma^N e^{-2\eta} \\
		\sigma_{01}^*\Gamma^{*N} e^{-2\eta} & \sigma_{11}
	\end{pmatrix}.
\end{eqnarray}
Then, the QFI associated with $\rho_\mathpzc{A}(t_1)$ reads
\begin{eqnarray}\label{Example-QFI-SM}
	\mathcal{F}_N^\textrm{noise}(T)=
	4N^2\abs{\sigma_{01}}^2\abs{\Gamma}^{2N-2}\abs{\partial_{_T}\Gamma}^2e^{-4\eta}+
	4N^2\abs{\sigma_{01}}^4\abs{\Gamma}^{4N-2}
	\left(\partial_{_T}\abs{\Gamma}\right)^2e^{-8\eta}
	/\left(\sigma_{00}\sigma_{11}-\abs{\sigma_{01}}^2\abs{\Gamma}^{2N}e^{-4\eta}\right).
\end{eqnarray}
Notably, $\mathcal{F}_N^\textrm{noise}(T)$ depends on $\rho$ through $\Gamma$ which, as can be seen from Eq.~(\ref{nt3-Gamma}), is dependent on $\rho_{00}$ and $\rho_{11}$ and is independent of $\rho_{01}$.

We first examine the limiting case of $\xi\rightarrow\infty$. Using Eq.~(\ref{nt3-Gamma}), we have that $\lim_{\xi\rightarrow\infty}\Gamma=\exp(-2i\varphi_{_T})$, which further leads to 
\begin{eqnarray}\label{Example-QFI-limit}
	\lim_{\xi\rightarrow\infty}\mathcal{F}_N^\textrm{noise}(T)=
	16N^2\abs{\sigma_{01}}^2\abs{\partial_{_T}\varphi_{_T}}^2e^{-4\eta}.
\end{eqnarray}
With Eq.~(\ref{Example-QFI-limit}), we reach the following observations: (i) $\lim_{\xi\rightarrow\infty}\mathcal{F}_N^\textrm{noise}(T)$ is irrespective of the  state $\rho^{\otimes N}$ of the $N$ probes, and (ii) $\lim_{\xi\rightarrow\infty}\mathcal{F}_N^\textrm{noise}(T)$ is a quadratic function of the absolute value $\abs{\sigma_{01}}$ of the off-diagonal element  of the state $\sigma$ of the ancilla. Note that $\abs{\sigma_{01}}$ is proportional to the coherence of $\sigma$ quantified via the so-called $l_1$ norm of coherence \cite{2018Zhang170501}, 
\begin{eqnarray}
	C_{l_1}(\sigma)=\sum_{i\neq j}\abs{\sigma_{ij}}=2\abs{\sigma_{01}}.
\end{eqnarray}
So, observation (ii) amounts to the statement that $\lim_{\xi\rightarrow\infty}\mathcal{F}_N^\textrm{noise}(T)$ is a quadratic function of the $l_1$ norm of coherence  of $\sigma$. We expect that the above two observations 
approximately hold for a large $\xi$. To confirm this point, we numerically compute $\mathcal{F}_N^\textrm{noise}(T)$ in units of $\mathcal{F}_\textrm{th}(T)$ for $\xi=400$. The numerical results are presented in Supplementary Fig.~\ref{fig-initial-state}. In Supplementary Fig.~\ref{fig-initial-state}a,  $\rho$ is randomly chosen whereas $\sigma$ is set to  $(\ket{0}+\ket{1})(\bra{0}+\bra{1})/2$. As can be seen from this subfigure, $\mathcal{F}_N^\textrm{noise}(T)$ does not change much in the course of varying $\rho_{00}$, which confirms that observation (i) approximately hold. In Supplementary Fig.~\ref{fig-initial-state}b, $\sigma$ is randomly chosen whereas $\rho$ is set to the steady state $\rho_{_T}$. As can be seen from this subfigure, $\mathcal{F}_N^\textrm{noise}(T)$ grows quadratically  as a function of $\abs{\sigma_{01}}$, which confirms that observation (ii) approximately hold.

\begin{figure}
	\includegraphics[width=0.5\textwidth]{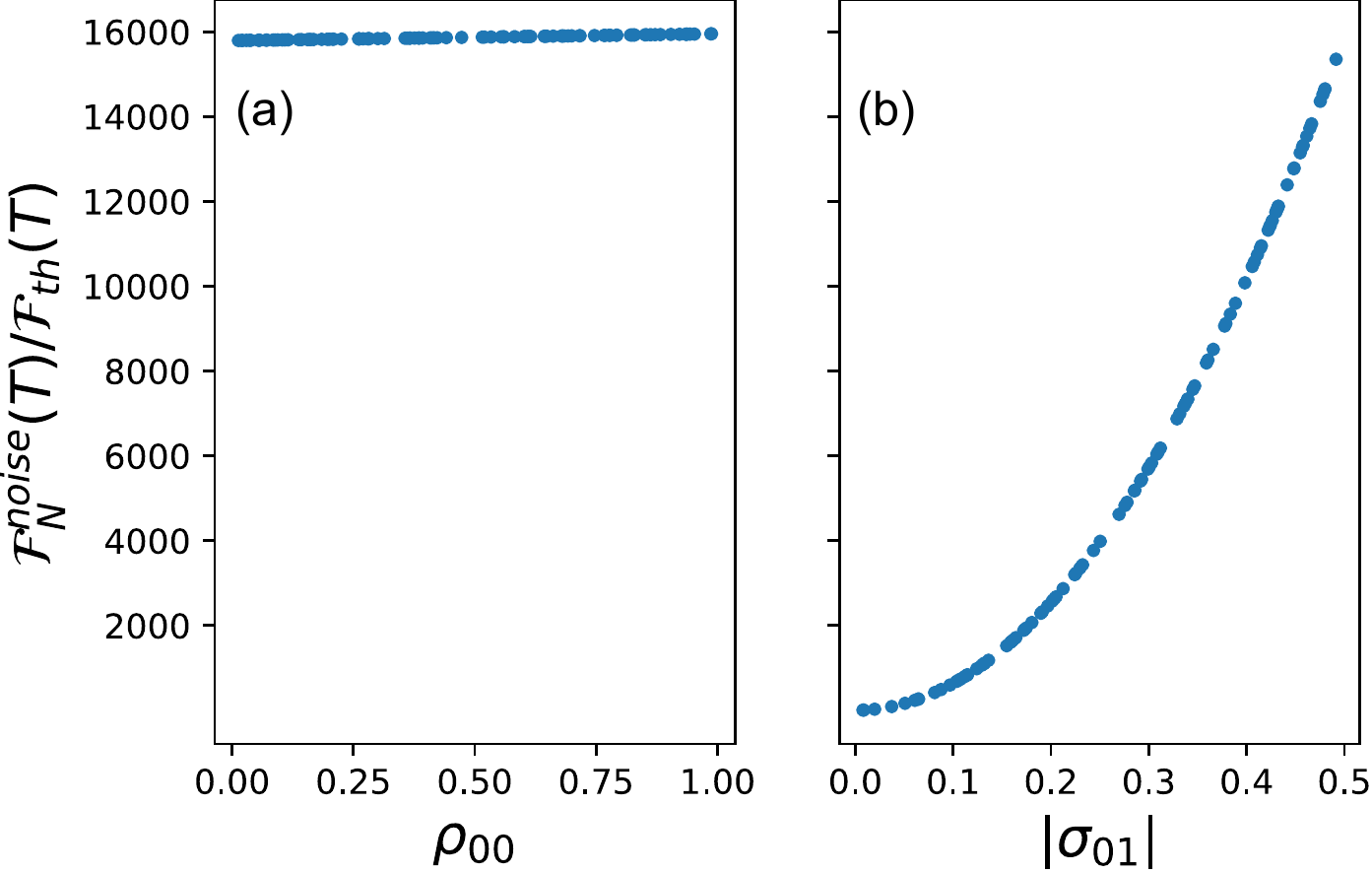}
	\caption{{Numerical results of $\mathcal{F}_N^\textrm{noise}(T)$ in units of $\mathcal{F}_\textrm{th}(T)$ for different initial states of the $N$ probes and the ancilla.} Here, the form of the initial state is assumed to be $\rho^{\otimes N}\otimes\sigma$. (a) $\rho$ is randomly chosen whereas $\sigma$ is set to  $(\ket{0}+\ket{1})(\bra{0}+\bra{1})/2$. (b) $\sigma$ is randomly chosen whereas $\rho$ is set to the steady state $\rho_{_T}$. Parameters used are $k_B T/\hbar\Omega=2$, $N=100$, $\xi=400$, and $\eta=1/10$. }
	\label{fig-initial-state}
\end{figure}

\section*{Supplementary Note 4: Proof of Eq.~(\inequalitysm) in the main text}

We first collect some useful facts:

(i) $\mathcal{L}_\mathpzc{S}\mathcal{P}=\mathcal{P}\mathcal{L}_\mathpzc{S}=0$. This follows from the facts $\mathcal{L}_\mathpzc{S}\rho_{_T}=0$ and $[\mathcal{L}_\mathpzc{S},\mathcal{P}]=0$.

(ii) $\mathcal{L}_\mathpzc{S}\mathcal{Q}=\mathcal{Q}\mathcal{L}_\mathpzc{S}=\mathcal{L}_\mathpzc{S}$.
This follows immediately from fact (i).


(iii) For any complex number $c$, operators $X$, $Y$, and superoperator $\mathcal{E}$, $\norm{cX}=\abs{c}\norm{X}$, $\norm{X+Y}\leq\norm{X}+\norm{Y}$, $\norm{XY}\leq\norm{X}\norm{Y}$, and $\norm{\mathcal{E}(X)}\leq\norm{\mathcal{E}}\norm{X}$.

(iv) $\mathcal{L}_\mathpzc{S}$ has a non-degenerate zero eigenvalue, and all its nonvanishing eigenvalues have negative real parts \cite{2016Zhang12117,2016Zhang52132}.

Let $\lambda_\mu$, $\mu=0,\cdots, M$, denote the eigenvalues of $\mathcal{L}_\mathpzc{S}$. We arrange them in ascending order of the absolute values of their real parts; that is, when $\mu <\nu$, there is $\abs{\Re(\lambda_\mu)}\leq \abs{\Re(\lambda_\nu)}$. So, $\lambda_0=0$, and $\abs{\Re(\lambda_\mu)}>0$, for $\mu=1,\cdots, M$.  $\lambda:={\min_{\mu>0}\abs{\Re(\lambda_\mu)}}=
{\abs{\Re(\lambda_1)}}$ is known as \textit{the dissipative gap}.
In the following, we prove the inequality (\inequalitysm) in the main text,
\begin{eqnarray}
	\norm{\mathcal{G}(t,s)\mathcal{Q}}\leq \varepsilon \exp[{\left(\varepsilon g\norm{\mathcal{K}}-\lambda\right)(t-s)}],
\end{eqnarray}
where $\varepsilon$ is a dimensionless constant determined by the damping basis of $\mathcal{L}_\mathpzc{S}$ [see Eq.~(\ref{varepsilon})], and $\mathcal{K}$ is the superoperator defined as $\mathcal{K}X:=-\textrm{i}[S\otimes A, X]$.

{First, we derive a convenient formula for $\mathcal{G}(t,s)\mathcal{Q}$.} Using fact (ii) and noting that $\mathcal{L}=\mathcal{L}_\mathpzc{S}+g\mathcal{K}$, we have $\mathcal{QL}=\mathcal{L}_\mathpzc{S}+g\mathcal{Q}\mathcal{K}$. It then follows that
\begin{eqnarray}\label{st1}
	\partial_t\mathcal{G}(t,s)=(\mathcal{L}_\mathpzc{S}+g\mathcal{Q}\mathcal{K})
	\mathcal{G}(t,s).
\end{eqnarray}
It is not difficult to see that Eq.~(\ref{st1}) can be formally solved as
\begin{eqnarray}\label{st2}
	\mathcal{G}(t,s)=e^{\mathcal{L}_\mathpzc{S}(t-s)}+g\int_{s}^{t}dt_1 e^{\mathcal{L}_\mathpzc{S}(t-t_1)}\mathcal{QK}\mathcal{G}(t_1,s).
\end{eqnarray}
Using Eq.~(\ref{st2}) iteratively, we obtain
\begin{eqnarray}
	\mathcal{G}(t,s)=e^{\mathcal{L}_\mathpzc{S}(t-s)}+g\int_{s}^{t}dt_1 e^{\mathcal{L}_\mathpzc{S}(t-t_1)}\mathcal{QK}e^{\mathcal{L}_\mathpzc{S}(t_1-s)}
	+g^2\int_{s}^{t}dt_1\int_{s}^{t_1}dt_2 e^{\mathcal{L}_\mathpzc{S}(t-t_1)}\mathcal{QK}e^{\mathcal{L}_\mathpzc{S}(t_1-t_2)}\mathcal{QK}
	e^{\mathcal{L}_\mathpzc{S}(t_2-s)}+\cdots.\nonumber\\
\end{eqnarray}
As an immediate consequence, we arrive at the desired formula:
\begin{eqnarray}\label{sm-formula}
	\mathcal{G}(t,s)\mathcal{Q}&=&e^{\mathcal{L}_\mathpzc{S}(t-s)}\mathcal{Q}+g\int_{s}^{t}dt_1 e^{\mathcal{L}_\mathpzc{S}(t-t_1)}\mathcal{QK}e^{\mathcal{L}_\mathpzc{S}(t_1-s)}\mathcal{Q}
	\nonumber\\
	&+&g^2\int_{s}^{t}dt_1\int_{s}^{t_1}dt_2 e^{\mathcal{L}_\mathpzc{S}(t-t_1)}\mathcal{QK}e^{\mathcal{L}_\mathpzc{S}(t_1-t_2)}\mathcal{QK}
	e^{\mathcal{L}_\mathpzc{S}(t_2-s)}\mathcal{Q}+\cdots.
\end{eqnarray}

{Second, we show that there exists a constant $\varepsilon$ such that
	\begin{eqnarray}\label{unit-norm}
		\norm{e^{\mathcal{L}_\mathpzc{S}(t-s)}\mathcal{Q}}\leq \varepsilon e^{-\lambda(t-s)},
	\end{eqnarray}
	i.e., $\norm{e^{\mathcal{L}_\mathpzc{S}(t-s)}\mathcal{Q}X}\leq \varepsilon e^{-\lambda(t-s)}$, for any operator $X$ with $\norm{X}\leq 1$.} An easy way to see this might be to employ the notion of the damping basis introduced in Supplementary Ref.~\cite{1993Briegel3311}. A damping basis of $\mathcal{L}_\mathpzc{S}$ is a set of operators, $\{R_\mu,\mu=0, \cdots, M\}$, such that
\begin{eqnarray}\label{damping-basis}
	\mathcal{L}_\mathpzc{S}X=\sum_{\mu=0}^{M}\lambda_\mu R_\mu\tr(L_\mu^\dagger X).
\end{eqnarray}
Here, $L_\mu$, $\mu=0, \cdots, M$, constitute a dual basis satisfying $\tr(L_\mu^\dagger R_\nu)=\delta_{\mu\nu}$. Evidently, $\mathcal{L}_\mathpzc{S}R_\mu=\lambda_\mu R_\mu$, i.e., $R_\mu$ is the eigenvector of $\mathcal{L}_\mathpzc{S}$ corresponding to the eigenvalue $\lambda_\mu$. Particularly, $R_0$ is the eigenvector associated with $\lambda_0=0$, i.e., $R_0=c\rho_{_T}$ for a coefficient $c$. We then have $QR_0=0$. Using fact (ii), we have $\mathcal{L}_\mathpzc{S}\mathcal{Q}R_\mu=\mathcal{Q}\mathcal{L}_\mathpzc{S}R_\mu=\lambda_\mu \mathcal{Q}R_\mu$ and $\mathcal{L}_\mathpzc{S}\mathcal{Q}R_\mu=\mathcal{L}_\mathpzc{S}R_\mu=\lambda_\mu R_\mu$.
So, $\mathcal{Q}R_\mu=R_\mu$, for $\mu=1, \cdots, M$. On the other hand, from Eq.~(\ref{damping-basis}), it follows that
\begin{eqnarray}\label{st3}
	e^{\mathcal{L}_\mathpzc{S}(t-s)}X=\sum_{\mu=0}^{M}e^{\lambda_\mu(t-s)} R_\mu\tr(L_\mu^\dagger X).
\end{eqnarray}
Using fact (ii), Eq.~(\ref{st3}), $QR_0=0$, and $QR_\mu=R_\mu$, for $\mu=1, \cdots, M$, we have
\begin{eqnarray}
	e^{\mathcal{L}_\mathpzc{S}(t-s)}\mathcal{Q}X=\mathcal{Q}e^{\mathcal{L}_\mathpzc{S}(t-s)}
	X=\mathcal{Q}\sum_{\mu=0}^{M}e^{\lambda_\mu(t-s)} R_\mu\tr(L_\mu^\dagger X)=\sum_{\mu=1}^{M}e^{\lambda_\mu(t-s)} R_\mu\tr(L_\mu^\dagger X).
\end{eqnarray}
So,
\begin{eqnarray}\label{st4}
	\norm{e^{\mathcal{L}_\mathpzc{S}(t-s)}\mathcal{Q}X}
	=\abs{e^{\lambda_1(t-s)}}\norm{\sum_{\mu=1}^{M}e^{(\lambda_\mu-\lambda_1)(t-s)} R_\mu\tr(L_\mu^\dagger X)}=e^{-\lambda(t-s)}\norm{\sum_{\mu=1}^{M}e^{(\lambda_\mu-\lambda_1)(t-s)} R_\mu\tr(L_\mu^\dagger X)}.
\end{eqnarray}
Using fact (iii) and the Cauchy-Schwarz inequality $\abs{\tr(L_\mu^\dagger X)}\leq \norm{L_\mu}\norm{X}$ and noting that $\norm{X}\leq 1$ and $\abs{e^{(\lambda_\mu-\lambda_1)(t-s)}}\leq 1$, we have
\begin{eqnarray}\label{st5}
	\norm{\sum_{\mu=1}^{M}e^{(\lambda_\mu-\lambda_1)(t-s)} R_\mu\tr(L_\mu^\dagger X)}
	\leq \sum_{\mu=1}^M \abs{e^{(\lambda_\mu-\lambda_1)(t-s)}}\abs{\tr(L_\mu^\dagger X)}\norm{R_\mu}\leq \sum_{\mu=1}^M\norm{L_\mu}\norm{R_\mu}.
\end{eqnarray}
Combining Eqs.~(\ref{st4}) and (\ref{st5}), we obtain Eq.~(\ref{unit-norm}) with
\begin{eqnarray}\label{varepsilon}
	\varepsilon=\sum_{\mu=1}^M\norm{L_\mu}\norm{R_\mu}.
\end{eqnarray}

{Third, with the aid of Eqs.~(\ref{sm-formula}) and (\ref{unit-norm}), we evaluate $\norm{\mathcal{G}(t,s)\mathcal{Q}}$.} Using fact (iii) and Eqs.~(\ref{sm-formula}) and (\ref{unit-norm}), we have
\begin{eqnarray}\label{st6}
	\norm{\mathcal{G}(t,s)\mathcal{Q}}&\leq&
	\norm{e^{\mathcal{L}_\mathpzc{S}(t-s)}\mathcal{Q}}
	+g\int_{s}^{t}dt_1 \norm{e^{\mathcal{L}_\mathpzc{S}(t-t_1)}\mathcal{Q}}
	\norm{\mathcal{K}}\norm{e^{\mathcal{L}_\mathpzc{S}(t_1-s)}\mathcal{Q}}\nonumber\\
	&+&g^2\int_{s}^{t}dt_1\int_{s}^{t_1}dt_2 \norm{e^{\mathcal{L}_\mathpzc{S}(t-t_1)}\mathcal{Q}}
	\norm{\mathcal{K}}\norm{e^{\mathcal{L}_\mathpzc{S}(t_1-t_2)}\mathcal{Q}}\norm{\mathcal{K}}
	\norm{e^{\mathcal{L}_\mathpzc{S}(t_2-s)}\mathcal{Q}}+\cdots\nonumber\\
	&\leq&\varepsilon e^{-\lambda(t-s)}+\varepsilon^2g\norm{\mathcal{K}}(t-s)e^{-\lambda(t-s)}
	+\frac{1}{2!}\varepsilon^3g^2\norm{\mathcal{K}}^2(t-s)^2e^{-\lambda(t-s)}+\cdots\nonumber\\
	&=&\varepsilon \left[1+\varepsilon g\norm{\mathcal{K}}(t-s)+
	\frac{\varepsilon^2 g^2\norm{\mathcal{K}}^2(t-s)^2}{2!}+\cdots\right] e^{-\lambda(t-s)}\nonumber\\
	&=&\varepsilon\exp[{\left(\varepsilon g\norm{\mathcal{K}}-\lambda\right)(t-s)}].
\end{eqnarray}
This completes the proof of Eq.~(\inequalitysm) in the main text.

\end{document}